\documentclass[10pt,letterpaper]{article}
\pdfoutput=1

\usepackage{jcappub}
\usepackage{amsmath,amssymb,color}
\usepackage{enumerate,xstring}
\usepackage{jcappub}
\usepackage{amsmath}
\usepackage{amsfonts}
\usepackage{amssymb}
\usepackage{graphicx}
\usepackage{epsfig}
\usepackage{color}
\usepackage{multirow}
\usepackage{graphicx}
\usepackage{bigints}
\usepackage{todonotes,bm,etoolbox}
\usepackage{calligra}
\usepackage{hyperref}
\usepackage{makecell}
\usepackage{tabularx}
\usepackage{booktabs}
\usepackage{subfigure}

\usepackage{tikz}

\def\ba#1\ea{\begin{align}#1\end{align}}
\def\bea{\begin{eqnarray}}
\def\eea{\end{eqnarray}}
\def\be{\begin{equation}}
\def\ee{\end{equation}}

\def\({\left(}
\def\){\right)}
\def\[{\left[}
\def\]{\right]}

\def\<{\left\langle}
\def\>{\right\rangle}

\def\comment#1{}

\def\eps{\epsilon}

\renewcommand{\v}[1]{\bm{#1}}

\def\vx{\v{x}}
\def\vk{\v{k}}
\def\vq{{\v{q}}}

\def\vtheta{\v{\theta}}

\def\vM{\v{M}}
\def\vD{\v{D}}

\newcommand{\perm}[1]{ \expandafter\ifstrempty\expandafter{#1} {\mbox{perm.}} {\mbox{$#1$ perm.}} }

\newcommand{\fnl}{f_\textnormal{\textsc{nl}}}

\newcommand{\bphi}{b_\phi}
\newcommand{\bphidelta}{b_{\phi\delta}}

\def\pathtofigs{./}

\definecolor{RedWine}{rgb}{0.743,0,0}
\definecolor{RoyalBlue}{rgb}{0.25,.41,.88}
\definecolor{ForestGreen}{rgb}{.13,.54,.13}
\definecolor{Goldenrod}{rgb}{.85,.65,.13}

\newcommand{\bq}{\begin{eqnarray}}
\newcommand{\eq}{\end{eqnarray}}

\title{Can we actually constrain $f_{\rm NL}$ using the scale-dependent bias effect? \vspace{1mm} \newline \Large An illustration of the impact of galaxy bias uncertainties using the BOSS DR12 galaxy power spectrum}

\author[a,b]{Alexandre Barreira}

\affiliation[a]{\small Excellence Cluster ORIGINS, Boltzmannstra\ss e 2, 85748 Garching, Germany}
\affiliation[b]{\small Ludwig-Maximilians-Universit\"at, Schellingstra\ss e 4, 80799 M\"unchen, Germany}

\emailAdd{alex.barreira@origins-cluster.de}

\date{\today}

\abstract{The scale-dependent bias effect on the galaxy power spectrum is a very promising probe of the local primordial non-Gaussianity (PNG) parameter $f_{\rm NL}$, but the amplitude of the effect is proportional to $f_{\rm NL}b_{\phi}$, where $b_{\phi}$ is the linear PNG galaxy bias parameter. Our knowledge of $b_{\phi}$ is currently very limited, yet nearly all existing $f_{\rm NL}$ constraints and forecasts assume precise knowledge for it. Here, we use the BOSS DR12 galaxy power spectrum to illustrate how our uncertain knowledge of $b_{\phi}$ currently prevents us from constraining $f_{\rm NL}$ with a given statistical precision $\sigma_{f_{\rm NL}}$. Assuming different fixed choices for the relation between $b_{\phi}$ and the linear density bias $b_1$, we find that $\sigma_{f_{\rm NL}}$ can vary by as much as an order of magnitude. Our strongest bound is $f_{\rm NL} = 16 \pm 16\ (1\sigma)$, while the loosest is $f_{\rm NL} = 230 \pm 226\ (1\sigma)$ for the same BOSS data. The impact of $b_{\phi}$ can be especially pronounced because it can be close to zero. We also show how marginalizing over $b_{\phi}$ with wide priors is not conservative, and leads in fact to biased constraints through parameter space projection effects. Independently of galaxy bias assumptions, the scale-dependent bias effect can only be used to detect $f_{\rm NL} \neq 0$ by constraining the product $f_{\rm NL}b_{\phi}$, but the error bar $\sigma_{f_{\rm NL}}$ remains undetermined and the results cannot be compared with the CMB; we find $f_{\rm NL}b_{\phi} \neq 0$ with $1.6\sigma$ significance. We also comment on why these issues are important for analyses with the galaxy bispectrum. Our results strongly motivate simulation-based research programs aimed at robust theoretical priors for the $b_{\phi}$ parameter, without which we may never be able to competitively constrain $f_{\rm NL}$ using galaxy data.}

\begin{document}

\maketitle

% ============================================================================================== %
% ============================================================================================== %
\section{Introduction}
\label{sec:intro}

Observational searches for local-type primordial non-Gaussianity (PNG) offer one of the most interesting ways to shed light on the physics behind the primordial density fluctuations generated during inflation. The level of local PNG is popularly parametrized by the parameter $\fnl$ in the equation \cite{2001PhRvD..63f3002K}
\bq\label{eq:fnl}
\phi(\vx) = \phi_{\rm G}(\vx) + \fnl\left[\phi_{\rm G}(\vx)^2 - \left<\phi_{\rm G}(\vx)^2\right>\right],
\eq
where $\phi(\vx)$ denotes the primordial gravitational potential and $\phi_{\rm G}$ is a Gaussian distributed random field. The simplest single-field models of inflation predict vanishing $\fnl$, and so a detection of $\fnl \neq 0$ would have very profound consequences in that it would imply multiple fields were present during inflation and the early Universe was not as simple as it could have been \cite{maldacena:2003, 2004JCAP...10..006C, 2011JCAP...11..038C, Tanaka:2011aj, conformalfermi, CFCpaper2, 2015JCAP...10..024D}. The current tightest bounds on $\fnl$ come from the analysis of three-point statistics of the cosmic microwave background (CMB) by the Planck satellite, which constrain  $\fnl = -0.9 \pm 5.1\ (1\sigma)$ \cite{2020A&A...641A...9P}. The next improvements over this bound are expected to come from analyses of the late-time spatial distribution of galaxies \cite{2012MNRAS.422.2854G, 2014arXiv1412.4872D, 2014arXiv1412.4671A, 2015ApJ...814..145A, 2015PhRvD..92f3525A, 2016JCAP...05..009R, 2017PhRvD..95l3513D, 2018MNRAS.478.1341K, 2019Galax...7...71B, 2019ApJ...872..126M, 2019MNRAS.489.1950B, 2021JCAP...04..013M, 2021arXiv210609713S,2022arXiv220307506F}, and it has been claimed that future galaxy surveys have in principle the potential to probe $\fnl$ with order unity precision, $\sigma_{\fnl} \sim 1$. Reaching for the $\sigma_{\fnl} = 1$ mark has since become a major science goal in galaxy survey analyses, as even if this happens without a detection of $\fnl$ (i.e.~$|\fnl| < 1$), this will still allow to disfavour several inflation models that typically predict $\fnl \sim \mathcal{O}(1)$ \cite{2014arXiv1412.4671A, 2022arXiv220308128A}.

The most renowned observational imprint of $\fnl$ on the galaxy distribution comes in the form of a series of new {\it bias parameter} terms that are proportional to $\fnl$, and which have a distinctive scale-dependence that allows them to be distinguished from other physical contributions \cite{slosar/etal:2008, mcdonald:2008, giannantonio/porciani:2010, 2011JCAP...04..006B, assassi/baumann/schmidt} (see also Sec.~7 of Ref.~\cite{biasreview} for a review of galaxy bias and PNG). This signature was first discovered in the power spectrum of dark matter halos in $N$-body simulations in the seminal work of Ref.~\cite{dalal/etal:2008}, where this effect was coined {\it scale-dependent bias}. The tightest constraints on $\fnl$ using this effect reported to date were obtained using the power spectrum of quasar samples in the eBOSS survey, $\fnl = -12 \pm 21\ (1\sigma)$ \cite{2021arXiv210613725M} (see also Ref.~\cite{2019JCAP...09..010C}). More recently, Refs.~\cite{2022arXiv220111518D, 2022arXiv220401781C} have utilized the galaxy bispectrum of BOSS DR12 galaxies to constrain $\fnl = -30 \pm 29\ (1\sigma)$ and  $\fnl = -33 \pm 28\ (1\sigma)$, respectively. The precision of these constraints is still a factor of $\approx 5$ worse than that of the CMB (see Refs.~\cite{slosar/etal:2008, 2011JCAP...08..033X, 2013MNRAS.428.1116R, 2014PhRvD..89b3511G, 2014PhRvL.113v1301L, 2014MNRAS.441L..16G, 2015JCAP...05..040H} for past constraints on $\fnl$ using galaxy data), but as next-generation surveys probe larger volumes of the Universe, the statistical power of similar analyses may approach $\sigma_{\fnl} \sim 1$.

The physical origin of the scale-dependent bias effect is important to understand, and is as follows. In the presence of local PNG, the galaxy density contrast $\delta_g(\vx, z) = n_g(\vx, z)/\bar{n}_g(z) - 1$ can be expanded linearly in Fourier space as \cite{slosar/etal:2008, mcdonald:2008, giannantonio/porciani:2010, 2011JCAP...04..006B, assassi/baumann/schmidt}
\bq\label{eq:biasintro1}
\delta_g(\vk, z) = b_1(z)\delta_m(\vk, z) + \bphi(z)\fnl\phi(\vk) + \eps(\vk, z),
\eq
where $\eps$ is a stochastic/noise variable, $\delta_m$ is the matter density contrast, and $b_1 = {\rm dln}n_g/{{\rm d}}\delta_m$ and $\bphi = {\rm dln}n_g/{\rm d}(\fnl\phi)$ are two galaxy bias parameters defined, respectively, as the response of the number density of galaxies to mass density $\delta_m$ and primordial potential perturbations $\phi$. These two perturbation types are related as $\delta_m(\vk,z) = \mathcal{M}(k,z)\phi(\vk)$, where $\mathcal{M}(k,z) = (2/3)k^2T_m(k)D_{\rm md}(z)/(\Omega_{m0}H_0^2)$, with $T_m(k)$ the matter transfer function and $D_{\rm md}(z)$ the linear growth function normalized to the scale factor $a = (1+z)^{-1}$ during matter domination. On large scales, the matter transfer function goes to unity, $T_m(k) \to 1$, and Eq.~(\ref{eq:biasintro1}) can be written as
\bq\label{eq:biasintro2}
\delta_g(\vk, z) = \underbrace{\Bigg[b_1(z) + \frac{3\Omega_{m0}H_0^2}{2k^2D_{\rm md}(z)}\bphi(z)\fnl\Bigg]}_{k{\rm-dependent\ coefficient}}\delta_m(\vk, z) + \eps(\vk, z),
\eq
i.e., the coefficient that multiplies $\delta_m(\vk, z)$ is now a function of scale $k$, and this is what the authors of Ref.~\cite{dalal/etal:2008} called a scale-dependent bias. Strictly, this is a misnomer\footnote{This terminology is so widespread now that we continue to incurr on the misnomer here. Note also that in the literature this effect is often described as a simple promotion of $b_1$ to be a function of scale $b_1 \to b_1 + \Delta b(k)$, but this replacement is only valid at the power spectrum level \cite{jeong/komatsu:2009b}. For higher-order statistics, like the bispectrum, additional PNG bias parameters beyond $\bphi$ need to be taken into account.} since what is scale-dependent is not any of the bias parameters $b_1$ or $\bphi$, but the relation between the perturbations $\delta_m$ and $\phi$ associated with them. The power spectrum of Eq.~(\ref{eq:biasintro2}) thus results (in addition to the power spectrum of the noise term) in three terms 
\bq\label{eq:pkterms}
b_1^2P_{mm}\ \ \ \ \ \ ;\ \ \ \ \ \ \propto\bphi\fnl \frac{P_{mm}}{k^2}\ \ \ \ \ \ ;\ \ \ \ \ \ \propto(\bphi\fnl)^2 \frac{P_{mm}}{k^4},
\eq
where $P_{mm}$ is the matter power spectrum. The last two terms are the scale-dependent signatures that local PNG leaves on the observed large-scale galaxy power spectrum and that can in principle be used to constrain $\fnl$. 

The problem we discuss in this paper is related to the fact that the amplitude of these scale-dependent terms is proportional not only to the parameter $\fnl$ that we wish to constrain, but also the galaxy bias parameter $\bphi$. That is, {\it in order to constrain $\fnl$ using the scale-dependent bias effect, we need to make assumptions about the galaxy bias parameter $\bphi$, and therefore, assumptions about galaxy formation and evolution.} As noted above, the parameter $\bphi$ is defined as the response of galaxy formation to primordial potential perturbations in local PNG cosmologies, i.e., it describes the excess number of galaxies that form inside primordial $\phi$ perturbations, compared to cosmic mean. The parameter $\bphi$ is in general a function of redshift, galaxy formation physics and properties of the observed galaxies like their total mass, stellar mass, star formation rate, color, etc. It is therefore currently very uncertain, which poses a problem to local PNG searches since its effect in the $k$-dependent coefficient in Eq.~(\ref{eq:biasintro2}) is indistinguishable from that of $\fnl$.

In attempting to circumvent this problem, most of the works in the literature (inspired by the original approach of Ref.~\cite{dalal/etal:2008}) assume that there is a fixed, tight relation between the parameters $\bphi$ and $b_1$. The idea is that the parameter $b_1$ can be fitted for using the smaller-scale part of the power spectrum (higher-$k$, where $\fnl$ constributes negligibly), which fixes $\bphi$ through the assumed $\bphi(b_1)$ relation. This breaks the degeneracy between $\bphi$ and $\fnl$, allowing the latter to be constrained. The most popular realization of this strategy utilizes the bias parameter relation obtained assuming universality of the halo mass function, $\bphi(b_1) = 2\delta_c(b_1 - p)$, with $p = 1$ and where $\delta_c \approx 1.686$ is the critical density for spherical collapse. This relation is not a perfect description even for halos in gravity-only simulations \cite{grossi/etal:2009, desjacques/seljak/iliev:2009, 2010MNRAS.402..191P, 2010JCAP...07..013R, 2011PhRvD..84h3509H, scoccimarro/etal:2012, 2012JCAP...03..002W, baldauf/etal:2015, 2017MNRAS.468.3277B, 2020JCAP...12..013B}, and it is also not expected to hold for real-life tracers (despite its widespread adoption). The observational constraints using quasars quote also bounds on $\fnl$ assuming the same functional form, but with a different value of $p = 1.6$ \cite{slosar/etal:2008, 2019JCAP...09..010C, 2021arXiv210613725M}. This follows from assuming that all of the observed quasars live in halos that have just recently merged \cite{slosar/etal:2008}, which is also an idealized assumption. Further, Ref.~\cite{2022arXiv220401781C} utilized $p = 0.55$ for BOSS DR12 galaxies, inspired by the results from Refs.~\cite{2020JCAP...12..013B, 2022JCAP...01..033B} using galaxy formation simulations. These were obtained for galaxies in stellar mass bins in the IllustrisTNG galaxy formation model, but the extent to which this actually describes BOSS DR12 galaxies has never been verified. The point to note is that there is currently a large theory error on the $\bphi(b_1)$ relation of observed galaxies, which poses a serious problem to $\fnl$ inference analyses since, because of the perfect degeneracy with $\bphi$, wrong assumptions about it translate directly into wrong constraints on $\fnl$.

In Ref.~\cite{2020JCAP...12..031B}, we presented a focused discussion on the impact that $\bphi$ uncertainties have on the resulting $\fnl$ constraints (see also Sec.~4 of Ref~\cite{2022JCAP...01..033B}). This was done in the context of an idealized simulated likelihood analysis for a fictitious (but realistic) survey with a mock {\it multitracer} data vector generated directly from a specified theory model. In this paper, we continue this discussion by extending it to the case of real galaxy observations using the power spectrum of BOSS DR12 galaxies. The main takeaway messages from our results are that (i) different, but currently equally plausible assumptions about the $\bphi(b_1)$ relation translate directly into different inferred precisions on $\fnl$; and (ii) contrary to what one might have naively expected, marginalizing over uncertainties on $\bphi(b_1)$ with large uninformative priors is not conservative and can bias the constraints on $\fnl$ through projection effects in the parameter space. 

In the absence of a robust knowledge of galaxy formation and the bias parameter $\bphi$, this implies that existing constraints and forecasts on $\sigma_{f_{\rm NL}}$ are currently subject to a large theory systematic error and should be interpreted carefully as a result. Independently of galaxy bias uncertainties, the scale-dependent bias effect can only be used to quote the {\it significance of detection} (SoD) of $\fnl \neq 0$ through constraints on the parameter combination $\fnl\bphi$, but in this case the value of $\sigma_{f_{\rm NL}}$ remains unknown and the constraints cannot be compared with the CMB bound. Our analysis in this paper can be regarded as an expanded discussion on the issue of $\bphi$ uncertainties of the analyses of Refs.~\cite{2022arXiv220111518D, 2022arXiv220401781C}, who recently used the same galaxy samples to constrain $\fnl$ using the galaxy power spectrum and bispectrum. We note further that the conclusions of this paper, despite focused on BOSS DR12 data, are important and hold generically to any attempt to constrain $\fnl$ using the scale-dependent bias effect, irrespective of the exact tracer sample considered (quasars, emission line galaxies, neutral Hydrogen, etc.). 

It should be noted that the uncertain value of $\bphi$ is an issue that has been mentioned in a number of past $\fnl$ works (e.g.~Refs.~\cite{slosar/etal:2008, 2010JCAP...07..013R, 2019JCAP...09..010C} for discussions about the $\bphi$ parameter for recent mergers and quasars), but which we find is not yet sufficiently appreciated in the general $\fnl$-related literature given the prominence of results still focused on the actual $\fnl$ bounds (which depend on $\bphi$), as opposed to the SoD of $\fnl \neq 0$. One of the goals of this paper is also to raise awareness for the need to improve on our knowledge of the $\bphi$ parameter, and in particular, to motivate the development of data- or simulation-based approaches to determine accurate and precise priors of $\bphi$ for real-life galaxy samples. Our ability to constrain the {\it actual numerical value} of $\fnl$ using the scale-dependent bias effect critically depends on our ability to determine these theory priors on $\bphi$.

The rest of this paper is organized as follows. In Sec.~\ref{sec:analysis}, we specify the details of our constraint analysis, including the redshift space galaxy power spectrum data vector, its covariance matrix and our theory model. We also validate our analysis choices on simulated mock galaxy samples. Our main $\fnl$ constraint results using the BOSS DR12 power spectrum are presented in Sec.~\ref{sec:results}, where we focus in particular on the strong impact that $\bphi$ uncertainties have on the final results. In Sec.~\ref{sec:bispectrum} we comment on analyses with the galaxy bispectrum, and discuss how even in this case there is a strong impact of PNG bias uncertainties. We summarize in Sec.~\ref{sec:summary}. In App.~\ref{app:triangle} we display a number of additional plots with one- and two-dimensional marginalized constraints on the free parameters of our theory model.

% ============================================================================================== %
% ============================================================================================== %
\section{Analysis specifications}
\label{sec:analysis}

In this section, we describe the data and the theory model that we use to constrain the local PNG signal. We also validate our analysis choices using mock galaxy data.

% ============================================================================================== %
% ============================================================================================== %
\subsection{Galaxy power spectrum data}
\label{sec:model}

As the data vector, we consider the multipoles of the redshift-space galaxy power spectrum measured by Ref.~\cite{2021PhRvD.103j3504P} (see also Refs.~\cite{2021PhRvD.104l3529P, 2022PhRvD.105d3517P}) for the galaxies in data release 12 (DR12) of the BOSS galaxy survey \cite{2017MNRAS.470.2617A}. A welcoming feature of these specific measurements compared to conventional approaches is that they can be readily compared against perturbation theory predictions without the need to first convolve with the survey window function (see Ref.~\cite{2021PhRvD.103j3504P} for the details). We consider the power spectrum measured for 4 galaxy samples: 2 high redshift samples with $z_3 = 0.61$ in the north and south galactic caps (dubbed NGCz3 and SGCz3), and 2 lower redshift samples with $z_1 = 0.38$ in the north and south galactic caps (dubbed NGCz1 and SGCz1). The north samples were observed on a wider area on the sky and contain therefore a larger number of galaxies. The volume and number of galaxies for the samples \{NGCz3, SGCz3, NGCz1, SGCz1\} are $V = \{2.80, 1.03, 1.46, 0.53\}{\rm Gpc^3}/h^3$ (for our fiducial value of $h$) and $N_g = \{435741, 158262, 429182, 174819\}$, respectively.

We consider the measurements of the redshift space monopole and quadrupole (cf.~Fig.~\ref{fig:bf} below). For the covariance matrix, we consider the estimate from 2048 {\it MultiDark-Patchy} mock galaxy samples ({\it Patchy} from hereon), which were constructed to resemble the clustering properties of BOSS DR12 galaxies \cite{2016MNRAS.456.4156K, 2016MNRAS.460.1173R}. The power spectrum multipoles of the mocks were measured using the same method of Ref.~\cite{2021PhRvD.103j3504P} to measure the BOSS DR12 data vector.\footnote{Concretely, we use the power spectrum data that is available at \url{https://github.com/oliverphilcox/Spectra-Without-Windows}. We are extremely thankful to Oliver Philcox for making these data publicly available!} Measurements of the hexadecapole also exist, but we do not consider them in our analysis as they do not depend on $\fnl$ to leading order.

In our constraint analyses below we will remain ultra conservative and consider only the data measured up to wavenumbers $k_{\rm max} = 0.05\ h/{\rm Mpc}$ to guarantee that linear theory models remain valid. The minimum wavenumber considered is $k_{\rm min} = 0.01 \ h/{\rm Mpc}$, yielding a total of 8 $k$ values for each multipole. For the 2 multipoles and 4 galaxy samples considered this yields a total data vector size of $N_d = 8 \times 2 \times 4 = 64$. Importantly, we do not attempt to model or mitigate the impact of observational systematic effects in the data that may contaminate the signal on large scales in a way that can affect $\fnl$ constraints \cite{2012MNRAS.424..564R, 2013PASP..125..705P, 2013MNRAS.435.1857L, 2019MNRAS.482..453K, 2021arXiv210613724R}. In this sense, our analysis here is idealized in that it does not marginalize over any systematic uncertainties (this is as in the local PNG analyses of Refs.~\cite{2022arXiv220111518D, 2022arXiv220401781C}).

% ============================================================================================== %
% ============================================================================================== %
\subsection{Theory model}
\label{sec:model}

Our analysis choices in this paper are motivated by aiming for the simplest possible setup that is able to retrieve meaningful constraints on the local PNG signal. We thus restrict ourselves to working with linear theory for the galaxy power spectrum, which is only valid on sufficiently large scales, but which as we will see is sufficient to constraint local PNG since the scale-dependent effect peaks precisely on the largest observable scales. Our analysis is in this sense similar to the local PNG constraints derived using quasars in the eBOSS survey \cite{2021arXiv210613725M, 2019JCAP...09..010C}, which were obtained assuming also linear theory.

Concretely, we use the following expression for the anisotropic galaxy power spectrum in redshift space (see e.g.~Refs.~\cite{2021JCAP...05..015M, 2022arXiv220111518D, 2022arXiv220401781C} for the expressions of the next-to-leading-order 1-loop power spectrum)
\bq\label{eq:modelPkmu}
P_{gg}(k,\mu) = \Bigg[\left(b_1 + f\mu^2\right)^2 + \frac{2\left(b_1 + f\mu^2\right)\bphi\fnl}{\mathcal{M}(k)} + \frac{(\bphi\fnl)^2}{\mathcal{M}(k)^2}\Bigg] P_{mm}(k) + \frac{\alpha_P}{\bar{n}_g},
\eq
where $f = {\rm dln}D/{\rm dln}a$ is the usual linear growth factor, $\mu$ is the cosine of the angle between the line-of-sight direction and the wavevector $\vk$,  and $\alpha_P$ is a parameter that quantifies departures of the (assumed constant) shot-noise from the Poisson expectation ($\alpha_P = 0$ corresponds to Poisson shot noise since this has been subtracted from the measurements); to ease the notation, we dropped the redshift dependence from this expression, which we leave implicit. The $\mu$ dependence of the galaxy power spectrum can be organized by expanding in multipoles as
\bq\label{eq:Pkmuexp}
P_{gg}(k,\mu) = \sum_{\ell} P_{gg}^{\ell}(k) L_{\ell}(\mu),
\eq
where $L_{\ell}$ are Legendre polynomials and the multipoles $P_{gg}^{\ell}(k)$ are given by
\bq\label{eq:Pkmulti}
P_{gg}^{\ell}(k) = \frac{2\ell+1}{2} \int_{-1}^1 {\rm d}\mu P_{gg}(k,\mu) L_{\ell}(\mu).
\eq
As noted above already, in this paper we will consider the monopole ($\ell=0$) and quadrupole ($\ell=2$). Except for the parameter $\fnl$, we keep all other cosmological parameters fixed in this paper. In particular, we take the best-fitting cosmological parameters from Ref.~\cite{2022PhRvD.105d3517P}, obtained from an inference analysis using a 1-loop power spectrum model for the same galaxy samples we consider in this paper (cf.~left-hand side of their table VII; Ref.~\cite{2022PhRvD.105d3517P} runs also constraints using the galaxy bispectrum, but we consider their power-spectrum-only results for better consistency with our analysis here): $\Omega_b h^2 = 0.02268$, $\Omega_c h^2 = 0.1218$, $h = 0.6778$, $\sigma_8 = 0.75$, $n_s = 0.9649$. We evaluate the matter power spectrum and transfer functions using the CAMB code \cite{camb}. We also skip considering so-called relativistic effects that contribute to the galaxy power spectrum with the same scale-dependence as $\fnl\bphi$, and could therefore in principle result in biased constraints on local PNG if unaccounted for \cite{2011JCAP...10..031B, 2010PhRvD..82h3508Y, challinor/lewis:2011, gaugePk, 2016JCAP...05..009R, 2015ApJ...814..145A, 2020MNRAS.499.2598W, 2021JCAP...11..010V, 2022JCAP...01..061C}. These terms are however expected to be relatively unimportant at the level of the BOSS survey volume (though they can become important in future surveys).

In part of our analysis, we will also run constraints with Gaussian priors on the parameters $b_1$, taken also from the analysis of Ref.~\cite{2022PhRvD.105d3517P} using the same galaxy data. We will do so in order to rescue back some of the constraining power that is lost by our choice of a very conservative $k_{\rm max} = 0.05\ h/{\rm Mpc}$, which does not allow the parameter $b_1$ to be as precisely constrained. Concretely, our results with Gaussian priors on $b_1$ utilize the following constraints from Table VII of Ref.~\cite{2022PhRvD.105d3517P}:
\bq\label{eq:b1priors}
b_1^{\rm NGCz3} &=& 2.288 \pm 0.15 \ ,\nonumber \\
b_1^{\rm SGCz3} &=& 2.449 \pm 0.145\ , \nonumber\\
b_1^{\rm NGCz1} &=& 2.172 \pm 0.13\ , \nonumber \\
b_1^{\rm SGCz1} &=& 2.209 \pm 0.14\ .
\eq
We stress that our adoption of priors on $b_1$ should not be confused as an addition of prior information on galaxy formation and bias to the analysis, but it should rather be regarded as a simple strategy to utilize information from $k > 0.05\ h/{\rm Mpc}$ that would improve the constraints on the $b_1$ parameter. Note that this is self-consistent since we keep the cosmology fixed to that of Ref.~\cite{2022PhRvD.105d3517P}, from where we take the priors on $b_1$ for the exact same galaxy samples. Should we have chosen a different value of $\sigma_8$, for example, then the adoption of these priors on $b_1$ would be manifestly inconsistent.

% ============================================================================================== %
% ============================================================================================== %
\subsection{The $\bphi(b_1)$ parametrization}
\label{sec:model}

In our model, each galaxy sample contributes with three additional free parameters, $\{b_1$, $\bphi$ and $\alpha_P\}$, yielding in general a total of $4 \times 3 + 1 = 13$ free parameters, including $\fnl$. In our analysis we will, however, follow the standard approach in the literature to assume a relation between $\bphi$ and $b_1$. We find it useful to recap here the origin behind different $\bphi(b_1)$ relations encountered in the literature:

\begin{itemize}

\item \underline{Universality}. Assuming universality of the halo mass function, it is possible to derive $\bphi(b_1) = 2\delta_c(b_1 - 1)$. This is by far the most widely adopted relation in the literature, although there is no compelling reason to expect this to hold for real tracers of the large-scale structure.

\item \underline{Dark matter halos.} It is well known, for example, that even the simpler case of halos in gravity-only simulations does not exactly satisfy the universality relation \cite{grossi/etal:2009, desjacques/seljak/iliev:2009, 2010MNRAS.402..191P, 2010JCAP...07..013R, 2011PhRvD..84h3509H, scoccimarro/etal:2012, 2012JCAP...03..002W, baldauf/etal:2015, 2017MNRAS.468.3277B, 2020JCAP...12..013B}. The halo $\bphi(b_1)$ relation is instead better described by $\bphi(b_1) = q \times 2\delta_c(b_1 - 1)$ with $q \approx 0.8$.

\item \underline{Recent mergers.} It was argued in Ref.~\cite{slosar/etal:2008} (see also Ref.~\cite{2010JCAP...07..013R}) that $\bphi = 2\delta_c(b_1- 1.6)$ is a better description of recently-formed halos that could be the typical hosts of quasars. This expression has been used in $\fnl$ constraints using quasars \cite{slosar/etal:2008, 2019JCAP...09..010C, 2021arXiv210613725M}, and these works do note that it yields different constraints relative to the universality relation. The validity of this relation for quasars has however never been checked with dedicated simulation work, and as a result, the $\bphi(b_1)$ relation of quasars and their actual constraining power on $f_{\rm NL}$ remains currently unclear.

\item \underline{IllustrisTNG galaxies.} Reference~\cite{2020JCAP...12..013B} found that stellar-mass selected galaxies in the IllustrisTNG galaxy formation model approximately satisfy $\bphi = 2\delta_c(b_1-0.55)$. This relation was recently assumed in Ref.~\cite{2022arXiv220401781C} to constrain $\fnl$ using BOSS DR12 galaxy data. Note however that the dependence of $\bphi(b_1)$ on the assumed galaxy feedback model is still unknown, as well as the exact links between the simulated and observed galaxy samples. Reference \cite{2022JCAP...01..033B} subsequently showed that other selection criteria (including galaxy color and black hole mass accretion rate) may not even admit a simple analytical redshift-independent description for the $\bphi(b_1)$ relation.

\item \underline{Neutral hydrogen (21cm) with IllustrisTNG.} Reference \cite{2022JCAP...04..057B} has shown using also the IllustrisTNG model that the $\bphi(b_1)$ relation of the neutral Hydrogen distribution is below the universality relation, with $\bphi(b_1) = 2\delta_c(b_1 - p)$, $p \in [1.1, 1.4]$ being a more accurate description. Again, the dependence of this result on the galaxy feedback model is still unknown. This relation has never been adopted in 21cm data forecasts on $\fnl$.

\end{itemize}
To simplify our analysis, we will assume that the relation $\bphi(b_1) = 2\delta_c(b_1 - p)$ is satisfied by all four BOSS DR12 samples, and will treat $p$ as a free parameter. We stress that by making this assumption our results are already optimistic about the impact of $\bphi$ uncertainties on the $\fnl$ constraints, compared to the more general approach of treating $\bphi$ as a free parameter (or assuming different values of $p$) for each galaxy sample. While it is reasonable to assume that the galaxy selection in the north and south galactic caps is not too dissimilar and thus satisfies similar $\bphi(b_1)$ relations, it is less clear whether a $\bphi(b_1)$ relation that holds at $z_1 = 0.38$ would also hold at $z_3 = 0.61$. This reduces the number of free parameters to a total of 10. In another part of our analysis where we the discuss the significance of detection by constraining $\fnl\bphi$, we will have instead a total of 12 parameters: $\{b_1, \alpha_P, \left[\fnl\bphi\right]\}$ for each of the 4 samples.

\begin{figure}
\centering
\includegraphics[width=0.8\textwidth]{\pathtofigs 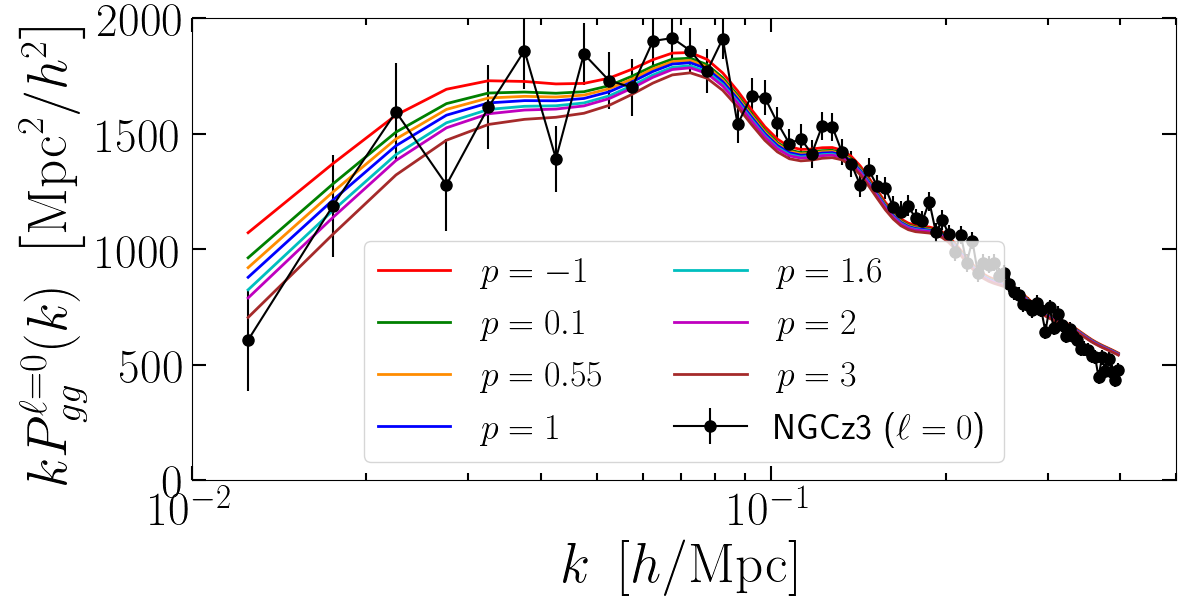}
\caption{Impact of different $\bphi(b_1)$ relations on the galaxy power spectrum. The result is shown for our linear theory model for $\fnl = 50$, $b_1 = 2.15$, $\alpha_P = 0$, and several values of the parameter $p \in \left[-1, 3\right]$ in the parametrization $\bphi(b_1) = 2\delta_c(b_1-p)$, as labeled. The black dots show the monopole of the galaxy power spectrum of the NGCz3 sample. The shape and amplitude of the $\bphi(b_1)$ relation is currently very uncertain, which translates directly into an uncertain impact of $\fnl$ on the large-scale galaxy power spectrum.}
\label{fig:p}
\end{figure}

Figure~\ref{fig:p} shows the monopole of the NGCz3 sample (black dots) together with the predictions of our linear theory model for different values of $p \in \left[-1, 3\right]$, as labeled. The result shown is for $\fnl = 50$, $b_1 = 2.15$ and $\alpha_P=0$. The figure makes apparent how different values of $p$ modify the amplitude of the signal on large scales, as thus how we can expect different constraints on $\fnl$ depending on our priors choices on $p$. The $\bphi(b_1)$ relation of the BOSS DR12 galaxies (or of any other tracer of the large-scale structure) is currently not known, which is precisely the issue that we discuss in this paper. Note also that this is a problem that cannot be resolved with the multitracer technique \cite{2009JCAP...10..007M, 2009PhRvL.102b1302S}, since each new galaxy sample that enters the multitracer analysis has its own associated $\bphi$ parameter, and the degeneracy with $\fnl$ remains unbroken \cite{2020JCAP...12..031B, 2022JCAP...01..033B}.

% ============================================================================================== %
% ============================================================================================== %
\subsection{Validation of the constraint analysis on the Patchy mocks}
\label{sec:mocks}

\begin{figure}
\centering
\includegraphics[width=\textwidth]{\pathtofigs 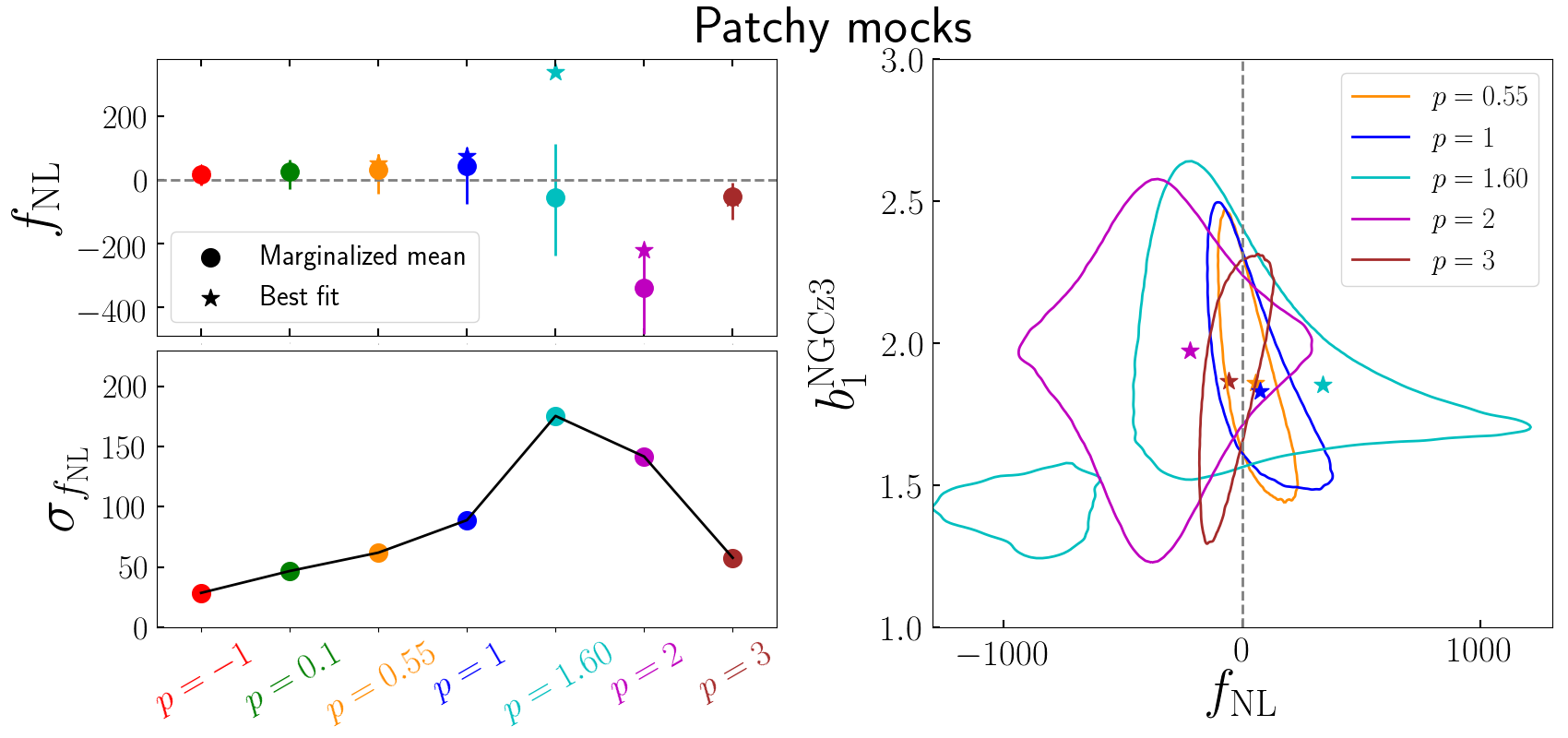}
\caption{Constraints on $\fnl$ from the Patchy mocks for different fixed $\bphi(b_1)$ relations. The upper left panel shows the marginalized one-dimensional $1\sigma$ constraints as a function of different fixed values of the $p$ parameter in the $\bphi = 2\delta_c(b_1-p)$ parametrization. The lower left panel plots $\sigma_{\fnl}$ vs.~$p$ to visualize better how different assumptions about the $\bphi(b_1)$ galaxy bias relation translate into different inferred precisions on $\fnl$. The right panel shows the two-dimensional $2\sigma$ marginalized constraints on the $b_1^{\rm NGCz3}-\fnl$ plane. The constraints are compatible with the fiducial value of $\fnl = 0$ of the mocks, which validates our analysis choices. The cases $p=1.6$ and $p=2$ are affected by projection effects, which is why the marginalized mean (circles) and maximum of the likelihood (stars) are offset (see the text for details).}
\label{fig:mocks}
\end{figure}

In order to validate our analysis choices, we run our constraint analysis taking as data vector the mean galaxy power spectrum multipoles of the 2048 Patchy mocks. Concretely, we constrain the parameter space using the following multivariate Gaussian likelihood function
\bq\label{eq:like}
-2{\rm ln}\mathcal{L}(\vtheta) = \left(\vD - \vM(\vtheta)\right)^t \hat{\v{C}}^{-1} \left(\vD - \vM(\vtheta)\right),
\eq
where $\vD$ is data vector, $\vM$ is the theory model prediction for a given set of model parameters $\vtheta$,  and $\hat{\v{C}}^{-1} = (N_r - N_d - 2) \v{C}^{-1} / (N_r - 1)$ is an unbiased estimate of the inverse covariance matrix \cite{2007A&A...464..399H} ($\v{C}^{-1}$ is the inverse of the covariance matrix $\v{C}$ obtained using a standard covariance estimator applied on the ensemble of $N_r = 2048$ realizations of the Patchy mocks). We sample the parameter space using the {\tt EMCEE} {\tt Python} implementation \cite{2013PASP..125..306F} of the affine-invariant Markov Chain Monte Carlo (MCMC) sampler in Ref.~\cite{2010CAMCS...5...65G}. We use 32 {\it walkers} and consider our chains to have converged when (i) the size of the chain is at least $100$ times the autocorrelation time and (ii) the latter has varied by less than $1\%$ since the last calculation point, which is every few thousand samples. We have also visually monitored the marginalized constraints during the course of the MCMC runs and found them to have become satisfactorily constant even before our nominal convergence criterion was reached. 

In this validation analysis, we run constraints for different fixed values of $p$ in the $\bphi(b_1) = 2\delta_c(b_1-p)$ parametrization, and sample the remainder of the 9-dimensional parameter space assuming wide, uninformative linear priors for all of the parameters. Further, in this validation part alone, we keep the cosmology fixed to that of the Patchy mocks (which is different from that we assume when we analyse the BOSS DR12 data): $\Omega_b h^2 = 0.02214$, $\Omega_c h^2 = 0.1189$, $h = 0.6777$, $\sigma_8 = 0.83$, $n_s = 0.9611$. The constraints on $\fnl$ from the Patchy mocks are displayed in Fig.~\ref{fig:mocks}. The upper left panel shows the one-dimensional $1\sigma$ marginalized constraints on $\fnl$ as a function of the fixed value of $p$. For most of the explored values of $p$, the fiducial value of $\fnl = 0$ of the mocks is recovered to within $1\sigma$. The noteworthy exception is the case with $p=2$ (magenta), whose constraint on $\fnl$ is $\approx 2.5\sigma$ below the true value. This is not a consequence of a breakdown of our analysis setup, but is rather due to projection effects that arise after marginalizing over poorly constrained directions in the parameter space. The galaxy samples in the Patchy mocks have values of $b_1 \approx 1.8 - 2$, and so when $p \approx 2$, the values of $\bphi \propto (b_1 - p)$ can become very small and thus the signal very insensitive to $\fnl$, which becomes poorly constrained. As an illustration, the right panel of Fig.~\ref{fig:mocks} shows the two-dimensional 2$\sigma$ marginalized constraints on the $b_1^{\rm NGCz3}-\fnl$ plane, where we note that the $p=2$ case comfortably brackets the true value of $\fnl=0$. Another indication that the $p=2$ constraints on $\fnl$ are prone to projection effects is the fact that the marginalized mean (circle) is far off from the maximum (unmarginalized) likelihood value (star). Note also that despite displaying unbiased $1\sigma$ constraints, the $p=1.6$ case (cyan) is also likely affected by projection effects.

The lower left panel of Fig.~\ref{fig:mocks} shows just the dependence of the inferred precision $\sigma_{\fnl}$ on the assumed value of the parameter $p$. For the values of $b_1\approx1.8-2$ that characterize the Patchy galaxy samples, $p = -1$ results in the largest values of $\bphi$, and thus in the tightest constraints on $\fnl$. The constraints become looser as $p$ increases to values comparable to the values of $b_1$, but they become tighter beyond that as $p$ increases to yield more negative values of $\bphi$ (a point beyond which the mean constraints on $\fnl$ switch sign). Indeed, as we have anticipated in our considerations above, the different values of $p$ result in different amplitudes for the galaxy bias parameter $\bphi$, which in turn result in different error bars on $\fnl$. We continue this discussion next using the BOSS DR12 data.

% ============================================================================================== %
% ============================================================================================== %
\section{Results from BOSS DR12}
\label{sec:results}

In this section we present and discuss our main results on the impact of the $\bphi$ parameter on local PNG constraints using the BOSS DR12 galaxy power spectrum. We discuss first in Sec.~\ref{sec:presults_pfixed} the impact of different fixed choices of the parameter $p$ in the parametrization $\bphi(b_1) = 2\delta_c(b_1-p)$, and then in Sec.~\ref{sec:presults_pprior} the impact of marginalizing over $p$. In Sec.~\ref{sec:sod} we show the constraints on the parameter combination $\fnl\bphi$, which is what can strictly be constrained by the data independently of prior assumptions on galaxy bias.

% ============================================================================================== %
% ============================================================================================== %
\subsection{Results from fixed $\bphi(b_1)$ relations}
\label{sec:presults_pfixed}

\begin{figure}
\centering
\includegraphics[width=\textwidth]{\pathtofigs 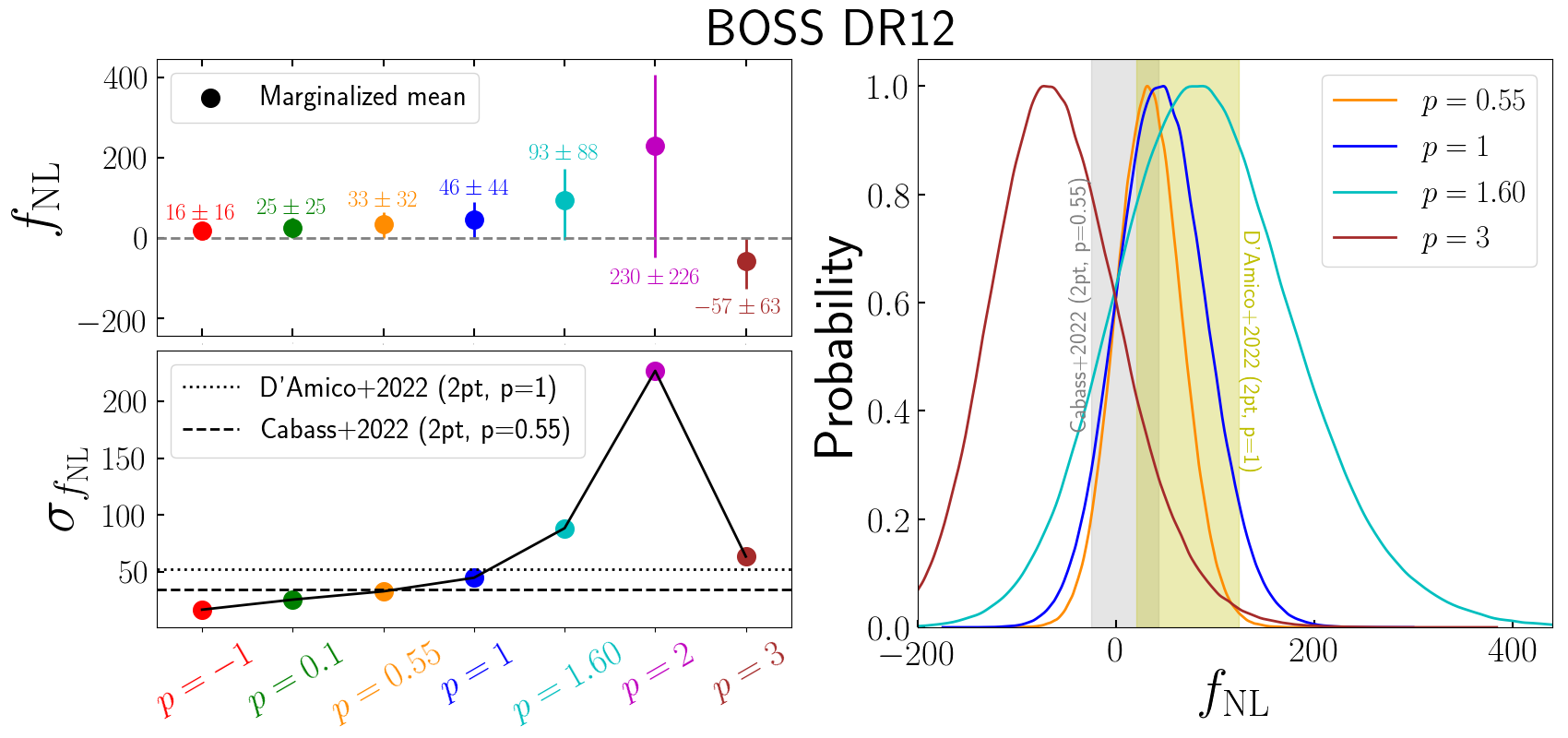}
\caption{Constraints on $\fnl$ using the BOSS DR12 galaxy power spectrum for different fixed $\bphi(b_1)$ relations. The result is shown for different values of $p$ in the parametrization $\bphi(b_1) = 2\delta_c(b_1 - p)$. The upper left panel shows the one-dimensional $1\sigma$ marginalized constraints, the lower left panel plots $\sigma_{\fnl}$ vs.~$p$, and the right panel shows the actual probability distribution for a few of the $p$ values. Marked also in the panels are the constraints obtained by Refs.~\cite{2022arXiv220111518D, 2022arXiv220401781C} using the same galaxies (we show only their power-spectrum-only results for comparison, but note they also utilize 3-point function information in their analysis). The takeaway message is that different assumptions on the uncertain value of $p$ result in different inferred precisions on $\fnl$.}
\label{fig:pfixed}
\end{figure}

\begin{table*}
\centering
\begin{tabular}{lcccccccccccccc}
\toprule
& $p$ value & $-1$ & $0.1$ & $0.55$ & $1$ &  $1.6$ & $2$ & $3$  \\
\midrule
\midrule
& No $b_1$ priors  & $47_{-38}^{+27}$ & $73_{-63}^{+42}$ & $94_{-85}^{+52}$ & $131_{-133}^{+63}$ &  $216_{-274}^{+84}$  &  $-338_{-627}^{+580}$ &  $-137_{-77}^{+44}$ \\
\midrule
& With $b_1$ priors  & $17_{-17}^{+16}$ & $26_{-26}^{+25}$ & $33_{-34}^{+32}$ & $46_{-46}^{+43}$ &  $93_{-97}^{+79}$  &  $230_{-278}^{+175}$ &  $-58_{-70}^{+57}$ \\
\bottomrule
\end{tabular}
\caption{Constraints on $\fnl$ using the BOSS DR12 galaxy power spectrum for different fixed $\bphi(b_1)$ relations. The result is for different values of $p$ in the parametrization $\bphi(b_1) = 2\delta_c(b_1 - p)$. The first line lists the constraints obtained without any priors on the bias parameters $b_1$, and the second line shows the result assuming the Gaussian priors of Eq.~(\ref{eq:b1priors}) from the results of Ref.~\cite{2022PhRvD.105d3517P}. The two approaches give compatible results, and both display the same dependence on the assumed value of $p$.}
\label{tab:pfixed}
\end{table*}

Figure \ref{fig:pfixed} shows the one-dimensional $1\sigma$ marginalized constraints on $\fnl$ obtained with the DR12 galaxy power spectrum, assuming different fixed values of the parameter $p$ in the parametrization $\bphi(b_1) = 2\delta_c(b_1 - p)$. Figure~\ref{fig:triangle_pfixed} in App.~\ref{app:triangle} shows a corner plot with the one- and two-dimensional marginalized constraints for the full parameter space. The result shown is for the analysis with Gaussian priors on the parameter $b_1$; Tab.~\ref{tab:pfixed} lists also the $\fnl$ constraints without these priors, which are consistent. We note that our conclusions on the impact of $\bphi$ uncertainties on $\fnl$ constraints do not depend on whether the priors on $b_1$ are assumed or not. In App.~\ref{app:triangle}, Fig.~\ref{fig:pfixed_nob1priors} shows the same as Fig.~\ref{fig:pfixed}, but without assumed priors on $b_1$: the numerical values of $\sigma_{\fnl}$ become larger, but importantly, the conclusion on the impact of $\bphi$ uncertainties  remains the same.

As we have anticipated from our validation analysis using the Patchy mocks in the previous section, Fig.~\ref{fig:pfixed} shows explicitly again that different choices for $p$ translate directly into different values of $\sigma_{\fnl}$. That is, the inferred precision on $\fnl$ is prior dominated and cannot be determined solely from the data. Concretely, the case with $p=-1$ is that which results in the largest numerical value of the PNG bias parameter ($\bphi \approx 10$ for $b_1 \approx 2$), and thus in the tightest constraints on $\fnl$, $\sigma_{\fnl} = 16$. As $p$ increases towards $p \to b_1 \approx 2$, $\bphi$ decreases towards $\bphi \to 0$, yielding looser and looser bounds on $\fnl$. Given our current limited knowledge of the $\bphi(b_1)$ relation, $p = 0.55$, $p = 1$, $p = 1.6$ and $p=2$ are all equally plausible options, but Fig.~\ref{fig:pfixed} and Tab.~\ref{tab:pfixed} show that $\sigma_{\fnl}$ can vary by a factor of $\approx 7$ within this range of $p$ values (and a factor of $\approx 14$ if we include the most extreme $p=-1$ case). Note that the impact of slightly different assumptions on $\bphi(b_1)$ can translate into substantially different constraints on $\fnl$ because the $\bphi(b_1)$ relation may cross zero. Finally, after $p$ crosses the typical values of $b_1$ of the BOSS DR12 galaxies, $\bphi$ increases again in absolute value (it becomes more negative; $\bphi \approx -3$ for $p = 3$), and the constraints on $\fnl$ begin to tighten up again.

The dotted and dashed horizontal lines in the lower left panel of Fig.~\ref{fig:pfixed} show the $\sigma_{\fnl}$ values from the power-spectrum-only part of the analyses of Ref.~\cite{2022arXiv220111518D} ($\sigma_{\fnl} = 52$) and Ref.~\cite{2022arXiv220401781C} ($\sigma_{\fnl} = 34$), respectively. These constraints are marked by the shaded vertical bands on the right panel. The bound from Ref.~\cite{2022arXiv220401781C} is a factor of $\approx 0.65$ smaller than that of Ref.~\cite{2022arXiv220111518D}, but this can be simply explained by the different values of $p$ assumed in the two analyses. Concretely, these results from Refs.~\cite{2022arXiv220111518D} and \cite{2022arXiv220401781C} are for $p = 1$ and $p=0.55$, respectively, which for $b_1 \approx 2$ means that Ref.~\cite{2022arXiv220111518D} has values of $\bphi$ that are $\approx 0.69$ smaller than those of Ref.~\cite{2022arXiv220401781C}, hence their correspondingly larger error bar. This is as one would expect from the perfect degeneracy between $\bphi$ and $\fnl$ in the galaxy power spectrum, and is again telling of how different assumptions on $\bphi(b_1)$ can have a sizeable impact on the inferred precision on $\fnl$.\footnote{Reference \cite{2022arXiv220401781C} also quotes constraints assuming $p=1$, in which case they find $\sigma_{\fnl} \approx 56$, as one would expect from the perfect degeneracy between $\bphi$ and $\fnl$. Likewise, in their constraints using eBOSS quasars, Refs.~\cite{2019JCAP...09..010C, 2021arXiv210613725M} quote constraints for $p=1$ and $p=1.6$, with the impact on $\sigma_{\fnl}$ being again as expected from the same degeneracy.} For comparison, these works report that the addition of the bispectrum information can lead to a reduction of $\sigma_{\fnl}$ of $20\%-40\%$, which is comparable to the difference in constraining power from two different, but currently plausible choices for $p$: $p = 0.55$ and $p=1$. Note also that for matching values of $p$, our $\sigma_{\fnl}$ values agree very well with those from Refs.~\cite{2022arXiv220111518D, 2022arXiv220401781C}.

An interesting result from Fig.~\ref{fig:pfixed} is also that, although different choices of $p$ result in different inferred precisions $\sigma_{\fnl}$, the {\it significance of detection} (SoD) of $\fnl \neq 0$ remains effectively unaffected. Concretely, the figure shows that the SoD is $\approx 1\sigma$ (consistent with no detection) for all values of $p$ shown. This is as expected from the perfect degeneracy between $\bphi$ and $\fnl$ at the power spectrum level. We note however that the robustness of the SoD to the exact value of $p$ should not be used as an argument to justify constraining local $\fnl$ in this way. For the sake of argument, consider the two observational bounds $\fnl^{\rm A} = 0.1 \pm 0.025$ and $\fnl^{\rm B} = 16 \pm 4$, which have the same SoD of $4\sigma$, and can both be obtained with the same galaxy data by making different assumptions about $\bphi$. These two bounds are however manifestly incompatible, i.e., erroneous assumptions about $\bphi$ introduce a theory systematic error that directly impacts the interpretation of the results. In this particular case, bound A is perfectly compatible with the Planck CMB constraint ($\fnl^{\rm CMB} = -0.9 \pm 5.1$), but bound B displays a $\gtrsim 3\sigma$ tension. The implications to models of inflation in order to generate predictions compatible with bound A or B would be also appreciably different. Concerning the SoD of $\fnl \neq 0$, the most transparent thing to do is to constrain the parameter combination $\fnl\bphi$ (cf.~Sec.~\ref{sec:sod} below).

Furthermore, should the real value for our Universe be $\fnl = 0$, then the precision $\sigma_{\fnl}$ is what is important to inform decisions about {\it when to stop} searching for local PNG and begin counting the {\it failed} search as evidence in favour of single-field inflation; for example, $|\fnl| < 0.025$ and $|\fnl| < 16$ are two constraints that can be possible with different choices of the $\bphi$ parameter, but which would be subject to very different interpretation. Note also that since the $\bphi(b_1)$ relation can cross zero, it is in fact not unrealistic that slightly different assumptions about $\bphi(b_1)$ can result in such different constraints on $\fnl$, as indeed shown by our results for $p = 1.6$ and $p=2$, for example.

% ============================================================================================== %
% ============================================================================================== %
\subsection{Results from marginalizing over the $\bphi(b_1)$ relation}
\label{sec:presults_pprior}

\begin{figure}
\centering
\includegraphics[width=\textwidth]{\pathtofigs 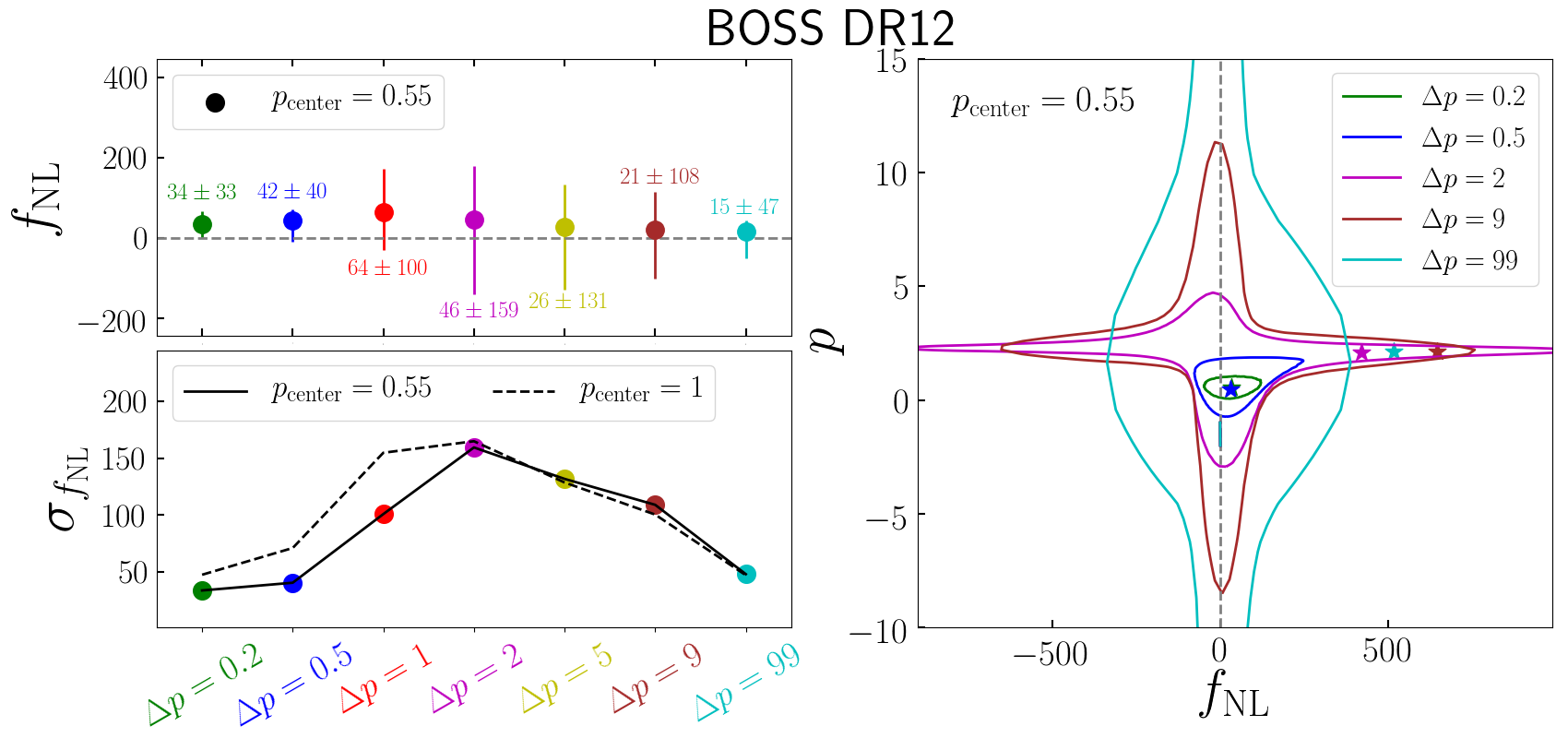}
\caption{Constraints on $\fnl$ using the BOSS DR12 galaxy power spectrum obtained by marginalizing over different assumed uncertainties on the $\bphi(b_1)$ relation. The result is shown for different values of the width $\Delta p$ of Gaussian priors on the parameter $p$ in the parametrization $\bphi(b_1) = 2\delta_c(b_1 - p)$; the Gaussian is assumed centered at $p_{\rm center} = 0.55$. The upper left panel shows the one-dimensional $1\sigma$ marginalized constraints as a function of $\Delta p$, and the lower left panel plots $\sigma_{\fnl}$ vs.~$\Delta p$; the lower left panel shows also the result for a prior centered at $p_{\rm center} = 1$ (dashed). The right panel shows two-dimensional $2\sigma$ marginalized constraints on the $p-\fnl$ plane for a number of $\Delta p$ values. The main takeaway is that there is no good choice for the $\bphi(b_1)$ prior that does not significantly inform the $\fnl$ constraints: (i) for lower $\Delta_p$, the constraints are impacted by the chosen values of $\Delta p$ and $p_{\rm center}$; and (ii) for larger $\Delta p$, the constraints cannot be trusted because they are severely impacted by projection effects that return artificially tight constraints around $\fnl = 0$.}
\label{fig:ppriors}
\end{figure}

\begin{table*}
\centering
\begin{tabular}{lcccccccccccccc}
\toprule
& $\Delta p$ value & $0.2$ & $0.5$ & $1$ & $2$ &  $5$ & $9$ & $99$  \\
\midrule
\midrule
& \makecell{ $p_{\rm center} = 0.55$ \\ (no $b_1$ priors) }  & $97_{-91}^{+50}$ & $86_{-85}^{+70}$ & $-29_{-268}^{+247}$ & $-123_{-163}^{+254}$ &  $-114_{-132}^{+201}$  &  $-106_{-110}^{+180}$ &  $-50_{-40}^{+96}$ \\
\midrule
& \makecell{ $p_{\rm center} = 0.55$ \\ (w/ $b_1$ priors) } & $34_{-35}^{+32}$ & $42_{-51}^{+29}$ & $64_{-94}^{+107}$ & $46_{-187}^{+131}$ &  $26_{-156}^{+107}$  &  $21_{-123}^{+94}$ &  $15_{-67}^{+29}$ \\
\midrule
& \makecell{ $p_{\rm center} = 1$ \\ (w/ $b_1$ priors) } & $49_{-51}^{+43}$ & $75_{-89}^{+53}$ & $60_{-152}^{+158}$ & $36_{-176}^{+154}$ &  $23_{-150}^{+107}$  &  $24_{-119}^{+81}$ &  $11_{-61}^{+32}$ \\
\bottomrule
\end{tabular}
\caption{Constraints on $\fnl$ using the BOSS DR12 galaxy power spectrum marginalizing over different assumed uncertainties on the $\bphi(b_1)$ relation. The result is for different values of the width $\Delta p$ of Gaussian priors on the parameter $p$ in the parametrization $\bphi(b_1) = 2\delta_c(b_1 - p)$. These results are for a Gaussian prior centered at $p_{\rm center} = 0.55$ (with and without $b_1$ priors) and $p_{\rm center} = 1$ (with $b_1$ priors).}
\label{tab:pprior}
\end{table*}

In cosmological inference analyses using galaxy data, our uncertain knowledge about galaxy formation is normally taken into account by marginalizing over the galaxy bias parameters. We will see next how this approach is ill-defined for the case of the $\fnl$ constraints because of projection effects associated with the degenerate nature of $\fnl$ and $\bphi$. In order to do so, we treat $p$ as a free parameter in our MCMC chains, and run constraints assuming different Gaussian priors on it
\bq\label{eq:pprior}
\mathcal{P}(p) \propto {\rm exp}\left[-\frac{1}{2}\frac{(p - p_{\rm center})^2}{\Delta p^2}\right].
\eq
Figure \ref{fig:ppriors} shows the constraints on $\fnl$ as a function of the prior width $\Delta p$; the main result is shown for priors centered at $p_{\rm center} = 0.55$, but the lower left panel displays also the result for $p_{\rm center} = 1$, as labeled. Table \ref{tab:pprior} displays the numerical values of the constraints. In the limit of $\Delta p \to 0$, we recover the same fixed-$p$ scenario discussed in the last section. The main result from Fig.~\ref{fig:ppriors} is that, while increasing the prior width up to $\Delta p \approx 2$ initially works to increase the error bar on $\fnl$, as the prior width increases beyond that, the error bar $\sigma_{\fnl}$ begins to shrink and the center value of the marginalized constraints becomes progressively centered around $\fnl = 0$. Concretely, for the $p_{\rm center} = 0.55$ case, from $\Delta p = 0.2$ to $\Delta p = 2$ the value of $\sigma_{\fnl}$ increases by a factor of $\approx 5$, but from $\Delta p = 2$ to $\Delta p = 5, 9$ and $99$, $\sigma_{\fnl}$ decreases by a factor of $\approx 1.2, 1.5$ and $3$, respectively.

This behavior has to do with marginalization projection effects, as illustrated in the right panel of Fig.~\ref{fig:ppriors} (see also Refs.~\cite{2020JCAP...12..031B, 2022JCAP...01..033B} for a more detailed explanation in the context of an idealized analysis for a fictitious survey). Concretely, along the $\fnl = 0$ direction, the parameter $p$ cannot be constrained since it only enters through terms that multiply $\fnl$. Thus, the wider the prior on $p$, the larger the fraction of the total parameter space volume that gets concentrated near $\fnl = 0$, which progressively biases the constraints after $p$ is marginalized over. This shows that, contrary to what one might have naively expected, wide priors on $p$ in particular, and in the $\bphi$ parameter in general, are not necessarily conservative and will still contribute to biased constraints on $\fnl$. In other words, given the perfect degeneracy between $\bphi$ and $\fnl$, there is no good choice for the size of the prior $\Delta p$ that does not significantly inform the constraints: small values of $\Delta p$ make the analysis sensitive to the center value of the prior $p_{\rm center}$, but large values of $\Delta p$ leave the analysis untrustworthy due to projection effects. 

In part of their analysis with the BOSS DR12 galaxies, the authors of Ref.~\cite{2022arXiv220111518D} have also explored the impact of marginalizing over the local PNG bias parameters. They considered a prior centered around the universality relation with a width given by $60\%$ of the value of the relation. This was done not only for the case of $\bphi$, but also the higher-order galaxy bias parameter $\bphidelta$ that enters their bispectrum analysis. There, this is reported to have resulted in an increase of $30\%$ of the error bar on $\fnl$. For the case of $\bphi$, and for the typical values of $b_1 \approx 2$ for the BOSS DR12 galaxies, the range of $\bphi$ values spanned by their $60\%$ uncertainty is equivalent to a case with $p_{\rm center} = 1$ and $\Delta p \approx 0.6$, in the $\bphi(b_1) = 2\delta_c(b_1-p)$ parametrization. Our closest scenario to this one in Fig.~\ref{fig:ppriors} is that with $p_{\rm center} = 1$ and $\Delta p = 0.5$, which causes the errorbar on $\sigma_{\fnl}$ to increase by $\approx 60\%$ relative to the $p = 1$ ($\Delta p = 0$) case in our fixed-$p$ analysis (cf.~Fig.~\ref{fig:pfixed}). We find this consistent with their degradation of $\approx 30\%$, noting that an exact match is not to be expected anyway given the differences between the two analyses (e.g.~Ref.~\cite{2022arXiv220111518D} utilizes also the bispectrum). In light of our discussion above, however, we expect that adopting wider priors in the analysis of Ref.~\cite{2022arXiv220111518D} will eventually begin returning biased constraints on $\fnl$ because of the projection effects.

In Ref.~\cite{2021JCAP...05..015M} the issue of marginalizing over $\bphi$ in $\fnl$ constraints has also been discussed in the context of simulated dark matter halo catalogues in real space. In the power-spectrum-only part of their analysis (but note Ref.~\cite{2021JCAP...05..015M} discusses also the halo bispectrum), the authors report constraints while fitting simultaneously for $\fnl$ and $\bphi$. In light of the perfect degeneracy between $\bphi$ and $\fnl$, this is only possible if priors on at least one of these parameters is assumed, but in which case the constraints become naturally dominated by the prior.\footnote{For completeness, we note that at the 1-loop level, there are contributions from $\fnl$ to the galaxy power spectrum that are not perfectly degenerate with $\bphi$. However, these are small and contribute with negligible constraining power.} For the specific case of an analysis of simulated data with a fiducial value of $\fnl = 10$, Ref.~\cite{2021JCAP...05..015M} reports that, compared to the case of assuming perfect knowledge of the $\bphi$ value of the halos (which can be known from separate universe simulations), marginalizing over a uniform prior with $\bphi \in \left[0, 6\right]$ increases the marginalized $2\sigma$ uncertainty on $\fnl$ from $\approx 10$ to $\approx 83$ (cf.~their Fig.~10). Note, however, that the latter bound is close to their assumed prior on $\fnl$, $\fnl \in \left[-100, 100\right]$, which is indicative that the result is prior dominated as one would expect. Again, in light of the projection effects discussed above, further increasing the width of the prior on $\bphi$ in Ref.~\cite{2021JCAP...05..015M} (including letting it explore negative values) would eventually bias the constraints towards $\fnl \to 0$ with a progressively smaller error bar. 

% ============================================================================================== %
% ============================================================================================== %
\subsection{Significance of detection analysis: constraints on ${\fnl}b_{\phi}$}
\label{sec:sod}

\begin{figure}
\begin{subfigure}
\centering
\includegraphics[width=0.5\textwidth]{\pathtofigs 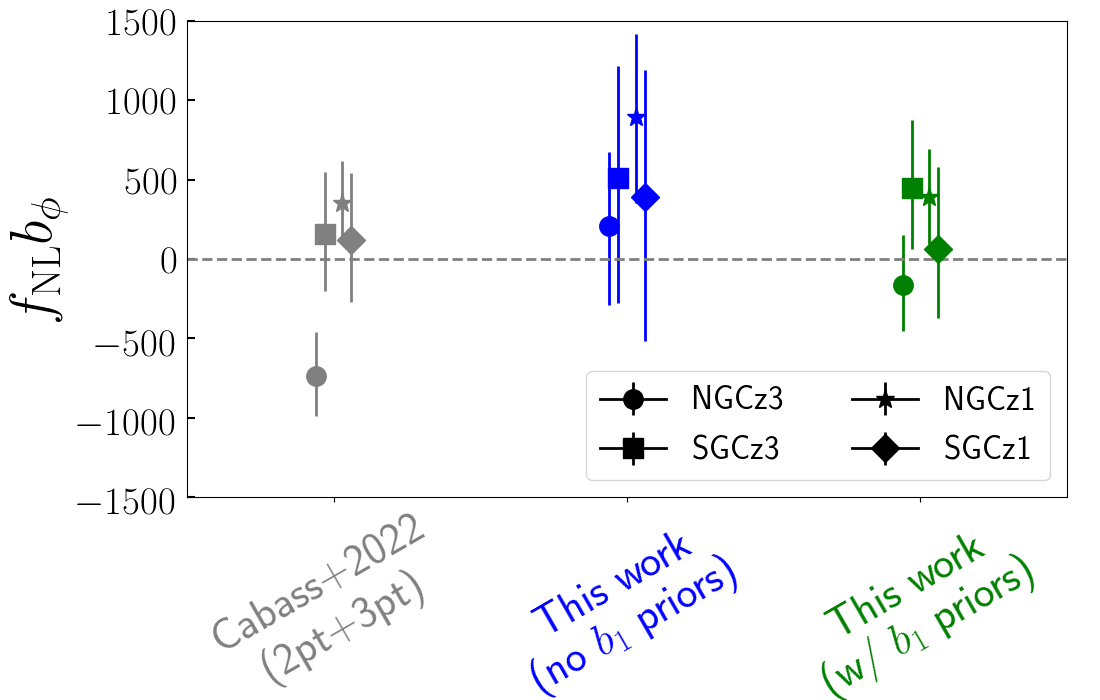}
\end{subfigure}
\begin{subfigure}
\centering
\includegraphics[width=0.5\textwidth]{\pathtofigs 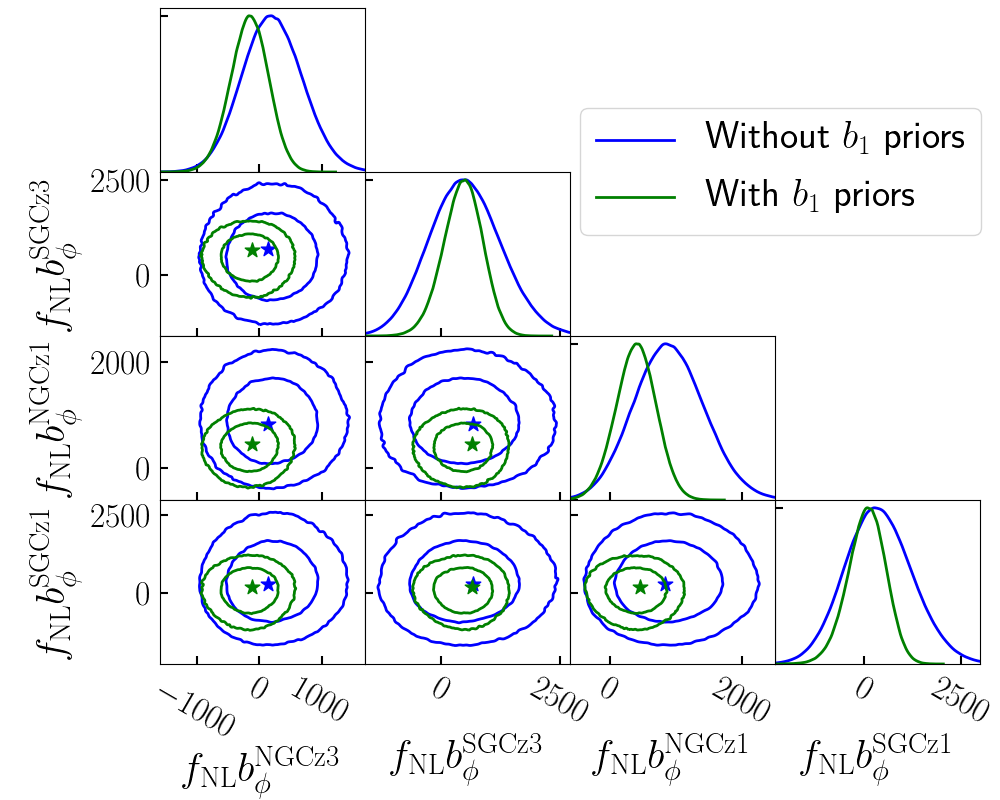}
\end{subfigure}
\caption{Constraints on the parameter combination $\fnl\bphi$ using the BOSS DR12 galaxy power spectrum. The left panel shows the one-dimensional $1\sigma$ marginalized constraints without (blue) and with (green) Gaussian priors on $b_1$. Shown also for comparison are the constraints obtained by Ref.~\cite{2022arXiv220401781C}, using also the bispectrum of the galaxies. The right panels show the two-dimensional marginalized $1\sigma$ and $2\sigma$ constraints on the subset of the parameter space made up of the $\fnl\bphi$ parameters of the four galaxy samples.}
\label{fig:bphifnl}
\end{figure}

\begin{table*}
\centering
\begin{tabular}{lcccccccccccccc}
\toprule
& Parameter & $\bphi\fnl^{\rm NGCz3}$ & $\bphi\fnl^{\rm SGCz3}$ & $\bphi\fnl^{\rm NGCz1}$ & $\bphi\fnl^{\rm SGCz1}$ \\
\midrule
\midrule
& No $b_1$ priors  & $210_{-499}^{+467}$ & $507_{-784}^{+709}$ & $898_{-542}^{+521}$ & $388_{-903}^{+806}$  \\
\midrule
& With $b_1$ priors  & $-163_{-289}^{+316}$ & $447_{-382}^{+428}$ & $391_{-295}^{+305}$ & $64_{-436}^{+518}$  \\
\bottomrule
\end{tabular}
\caption{Constraints on the parameter combination $\fnl\bphi$ using the BOSS DR12 galaxy power spectrum, with and without Gaussian priors on the parameter $b_1$.}
\label{tab:bphifnl}
\end{table*}

Independently of prior assumptions on $\bphi$ and using the scale-dependent bias effect, the galaxy power spectrum can only be used to constrain the parameter combination $\fnl\bphi$. Doing so does not let us constrain the numerical value of $\fnl$ and its uncertainty $\sigma_{\fnl}$, and it prevents also direct comparisons with the CMB data bounds. Note, however, that there is still value in placing constraints on $\fnl\bphi$ since they can still let us detect local PNG through detections of $\fnl\bphi \neq 0$ (under the only assumption that $\bphi \neq \infty$). These types of analyses are not yet routine in constraint/forecast works in the $\fnl$-related literature (see Refs.~\cite{2020JCAP...12..031B, 2022JCAP...01..033B} for the first discussions), but it is strongly recommended that this begins to be the case as this is what the scale-dependent bias effect can truly constrain.

\begin{figure}
\centering
\includegraphics[width=\textwidth]{\pathtofigs 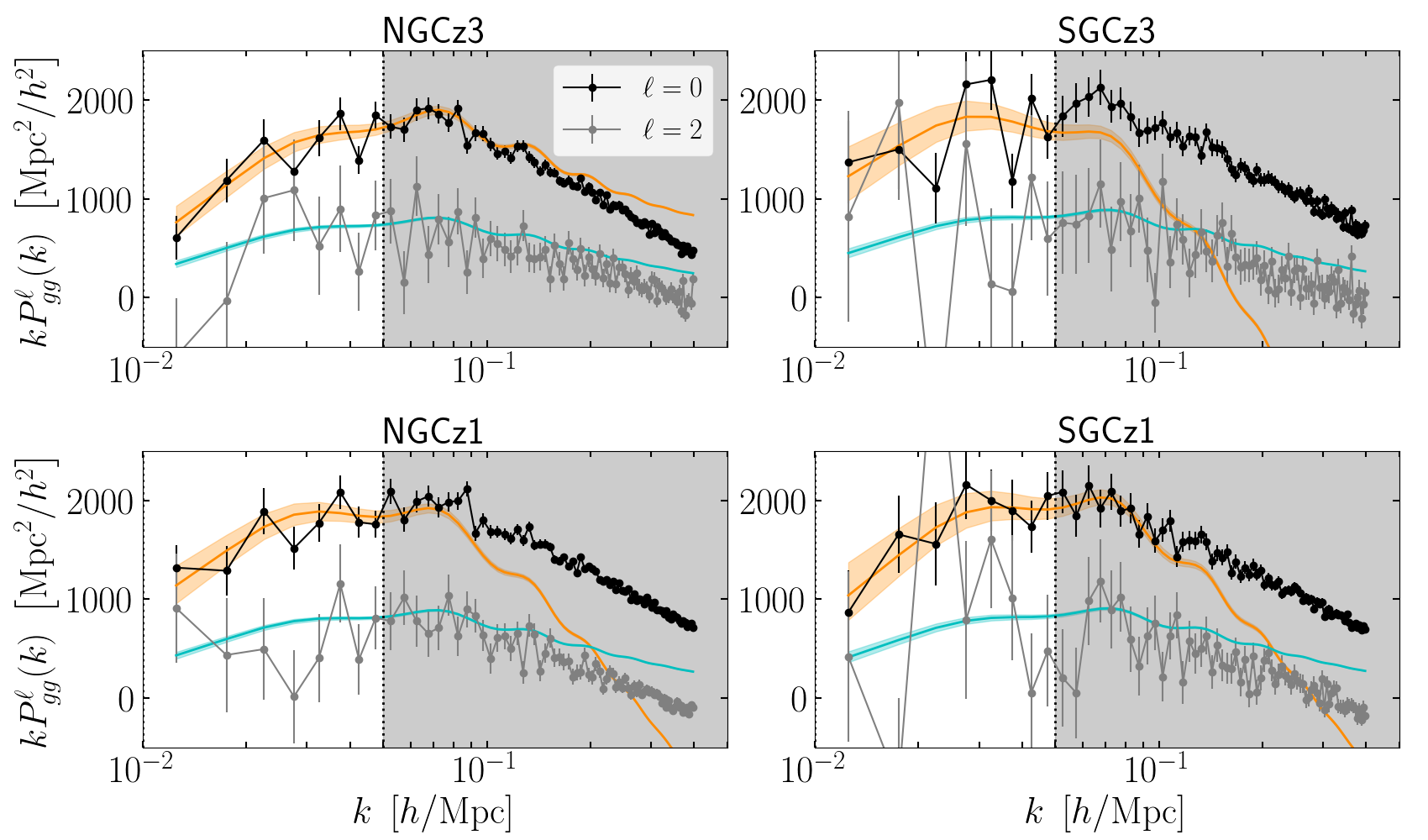}
\caption{The best-fitting model predictions obtained from the part of our analysis that fits for the parameter combination $\fnl\bphi$. The orange and cyan shaded bands mark the uncertainty on the monopole and quadrupole from the $1\sigma$ uncertainty on $\fnl\bphi$. Each of the four panels is for each of the galaxy samples considered; the black and grey dots with error bars show the measured monopole and quadrupole from Ref.~\cite{2021PhRvD.103j3504P}. The grey shaded bands mark the values of $k > k_{\rm max} = 0.05\ h/{\rm Mpc}$ that {\it were not} used to guarantee the validity of linear theory in our analysis.}
\label{fig:bf}
\end{figure}

Figure \ref{fig:bphifnl} and Tab.~\ref{tab:bphifnl} show our constraints on $\fnl\bphi$ from the analysis in which we treat this parameter combination as free for each of the four galaxy samples. The resulting best-fitting galaxy power spectrum is shown in Fig.~\ref{fig:bf}, and  Fig.~\ref{fig:triangle_fnlbphi} in App.~\ref{app:triangle} shows a corner plot with the one- and two-dimensional marginalized constraints for the full parameter space. In our analysis with (without) Gaussian priors on $b_1$, we recover inferred precisions on $\fnl\bphi$ of order $\sigma_{\fnl\bphi} \approx 300-500$ ($\sigma_{\fnl\bphi} \approx 450-900$). This is in line with the result from Ref.~\cite{2022arXiv220401781C} (grey points with error bars on the left of Fig.~\ref{fig:bphifnl}) who finds $\sigma_{\fnl\bphi} \approx 250-420$, using also information from the bispectrum. This shows that the power spectrum is what dominates the constraints on these SoD parameters, as first pointed out in Ref.~\cite{2020JCAP...12..031B}.

Assuming that the four $\fnl\bphi$ parameters are Gaussian distributed and independent (which is reasonable given the lack of any noticeable strong correlation on the right panels of Fig.~\ref{fig:bphifnl}), the overall SoD of $\fnl\bphi \neq 0$ (and thus $\fnl \neq 0$) is $\approx 1.6\sigma$ for both the case with and without $b_1$ priors (consistent with no overall detection of local PNG). Note that the SoD of $\approx 1\sigma$ in our fixed-$p$ analysis in Sec.~\ref{sec:presults_pfixed} (cf.~Fig.~\ref{fig:pfixed}) needs not to be the same as the SoD from the $\fnl\bphi$ analysis. This is because in the $\fnl\bphi$ case, the four galaxy samples contribute independently to the SoD (since we treat $\fnl\bphi$ as a free parameter for each sample), whereas in the fixed-$p$ analysis the different galaxy samples contribute with correlated information as we have assumed the same value of $p$ for all of them.

% ============================================================================================== %
% ============================================================================================== %
\section{Can the galaxy bispectrum help?}
\label{sec:bispectrum}

Relative to the power spectrum, the leading-order galaxy bispectrum receives two new types of contributions from local PNG that are interesting to discuss. The first comes from the primordial squeezed bispectrum signal that local PNG induces in the initial density perturbations. This signal is present also in the late-time matter bispectrum, and contributes to the galaxy bispectrum with an amplitude $\propto b_1^3 \fnl$, independently of any local PNG bias parameter. The second is the contribution from a new local PNG galaxy bias parameter $\bphidelta$ that contributes to the galaxy bias expansion as $\delta_g(\vx) \supset \bphidelta\fnl\delta_m(\vx)\phi(\vq(\vx))$ ($\vq$ is the Lagrangian coordinate of $\vx$). Schematically, the main $\fnl$ contributions in analyses with the tree-level galaxy bispectrum are (see e.g.~Refs.~\cite{giannantonio/porciani:2010, 2011JCAP...04..006B, sefusatti/etal:2012, assassi/baumann/schmidt, 2021JCAP...05..015M, 2022arXiv220111518D, 2022arXiv220401781C})
\bq\label{eq:Bgggterms}
\propto b_1^3\fnl\ \ \ \ \ \ \ \ ;\ \ \ \ \ \ \ \ \ \propto \fnl\bphidelta \ \ \ \ \ \ \ \ ; \ \ \ \ \ \ \ \ \propto \fnl\bphi
\eq
(we are dropping $\sim \fnl^2$ contributions for simplicity). The first point to note is that the new local PNG bias parameter $\bphidelta$ is currently also very uncertain, and so its contribution cannot be used to constrain $\fnl$ because of the same reasons discussed above for $\bphi\fnl$. Any constraint derived from this term must come from robust theoretical priors on $\bphidelta$, for example in the form of some $\bphidelta(b_1)$ relation, which are still lacking. This relation has only been recently measured for halos and IllustrisTNG galaxies in Ref.~\cite{2022JCAP...01..033B}, and for neutral Hydrogen in IllustrisTNG in Ref.~\cite{2022JCAP...04..057B}, but more work is needed to establish the robustness and connection of these early results to the real Universe.

An interesting question, however, concerns the ability of the galaxy bispectrum to constrain $\fnl$ through the primordial contribution $\propto b_1^3\fnl$, after marginalizing over the $\fnl\bphi$ and $\fnl\bphidelta$ parameters. This was first discussed in Refs.~\cite{2020JCAP...12..031B, 2022JCAP...01..033B} in the context of an idealized analysis for a data vector generated from a theory model. Unfortunately, there we found that the terms $\propto b_1^3 \fnl$ and $\propto \fnl\bphidelta$ have a very similar scale dependence, and thus that marginalizing over the parameter combination $\bphidelta\fnl$ severely washes out the contribution from the $b_1^3\fnl$ term, effectively resulting in very uncompetitive $\fnl$ bounds. This result was recovered by the more recent BOSS DR12 bispectrum analysis of Ref.~\cite{2022arXiv220401781C}, who also found a very strong degeneracy between $\fnl$ and $\fnl\bphidelta$ that keeps $\fnl$ from being constrained as competitively: compared to their nominal result of $\fnl = -33 \pm 28$ assuming fixed $\bphi(b_1)$ and $\bphidelta(b_1)$ relations, when $\fnl\bphi$ and $\fnl\bphidelta$ are marginalized over (including $\sim \fnl^2$ terms), the constraints degrade significantly to $\fnl = -676^{+150}_{-250}$.

At least to leading-order, this indicates that the galaxy bispectrum is unable to competitively constrain $\fnl$ independently of assumptions on the local PNG bias parameters. This is seemingly at odds with the results from Ref.~\cite{2021JCAP...05..015M} in their analysis of the bispectrum of dark matter halos in simulations with $\fnl \neq 0$. However, we note that although the authors of Ref.~\cite{2021JCAP...05..015M} quote factors of improvement from adding the bispectrum to analyses with the power spectrum when $\bphi$ and $\bphidelta$ are varied, the constraints on $\fnl$ in both cases are dominated by the assumed priors on $\bphi$ and $\bphidelta$ (as noted there). We note however that in this case one cannot refer to these priors as {\it loose} as they effectively control the resulting inferred precision on $\fnl$ (cf.~discussion in Sec.~\ref{sec:presults_pprior}). Again, in light of the strong impact of PNG bias uncertainties, a more transparent way to quote bounds on the local PNG signal is through limits on SoD parameter combinations like $\fnl\bphi$ and $\fnl\bphidelta$. 

In the future, it would be interesting to check the extent to which the situation changes in analyses with the 1-loop galaxy bispectrum \cite{2022arXiv220111518D}, or by probing higher-order statistics with the aid of field-level galaxy forward models \cite{2021arXiv211214645B, 2022arXiv220308838A} (though in Ref.~\cite{2022arXiv220308838A} the authors still assume perfect knowledge of the $\bphi(b_1)$ relation). We note however that if competitive constraints on $\fnl$ end up being possible with these higher-order analyses, these will come from our ability to probe the primordial signal, and not through the scale-dependent bias effect.

We note finally in passing that these considerations about the strong impact that PNG galaxy bias parameters have on $\fnl$ constraints apply primarily to the case of PNG of the {\it local-type}, and less so to the case of other shapes of PNG such as the {\it equilateral} and {\it orthogonal} shapes. Although for the equilateral and orthogonal cases there are also contributions that arise through new galaxy bias parameter terms, for order unity values of these bias parameters, these contributions are subdominant compared to the primordial signal imprinted in the matter bispectrum, which may be used to constrain $\fnl^{\rm equi.}$ and $\fnl^{\rm ortho.}$ relatively independently of PNG bias uncertainties \cite{2022arXiv220111518D, 2022arXiv220107238C}.

% ============================================================================================== %
% ============================================================================================== %
\section{Summary}
\label{sec:summary}

The determination of the numerical value of the local PNG parameter $\fnl$ would carry very profound consequences to our understanding of the early Universe and the primordial density fluctuations generated during the epoch of inflation.  In particular, detecting $\fnl \neq 0$ could be used to rule out the simplest single-field models of inflation in favor of more elaborate models involving multiple physical degrees of freedom. The current tightest bound comes from the Planck satellite CMB data analysis, which constrains $\fnl = - 0.9 \pm 5.1\ (1\sigma)$. This is compatible with single-field inflation, but leaves still significant room to accommodate several multi-field scenarios that predict order unity values. Reaching for the $\sigma_{\fnl} \lesssim 1$ mark has since become a major milestone in observational cosmology, as even if this happens without a clear detection of $\fnl \neq 0$, that can still be used to rule out many popular models of inflation. 

Currently, the best chances to reach for the $\sigma_{\fnl} \lesssim 1$ milestone are expected to come from constraints on the amplitude of the scale-dependent bias signatures that local PNG leaves on the statistics of the galaxy distribution. However, a problem that exists in these types of analyses is that the amplitude of these signatures is determined not only by the parameter $\fnl$, but also by a series of galaxy bias parameters that describe the response of galaxy formation to long-wavelength primordial gravitational potential perturbations. In the case of the galaxy power spectrum, the relevant bias parameter is called $\bphi$ and the constraints are dominated by contributions $\propto \fnl\bphi/k^2, (\fnl\bphi)^2/k^4$ (cf.~Eq.~(\ref{eq:modelPkmu})). Effectively all existing constraints on $\fnl$ reported to date using the scale-dependent bias effect rely on tight assumptions about $\bphi$, or more specifically its relation to the linear density bias parameter $b_1$. This is an issue to $\fnl$ constraints since the $\bphi(b_1)$ relation is currently still very uncertain, but its observational effects on the galaxy power spectrum are degenerate with $\fnl$.

In this paper we discussed the strong impact that the $\bphi(b_1)$ relation has on $\fnl$ constraints obtained using the scale-dependent bias effect. In particular, we considered the measurements of Ref.~\cite{2021PhRvD.103j3504P} of the redshift-space galaxy power spectrum of BOSS DR12 galaxies, and showed how our current uncertain knowledge of the $\bphi(b_1)$ relation prevents us from being able to determine $\fnl$ and its statistical uncertainty $\sigma_{\fnl}$. We adopted a simple and very conservative analysis setup assuming linear theory and considering only wavenumbers up to $k_{\rm max} = 0.05\ h/{\rm Mpc}$. We worked with the parametrization $\bphi(b_1) = 2\delta_c(b_1-p)$ that is often encountered in the literature, and constrained $\fnl$ assuming different fixed values of the parameter $p$, as well as marginalizing over its uncertain value. This specific parametrization of the $\bphi(b_1)$ relation served simply as a convenient way to parametrize our ignorance about the $\bphi$ parameter, and our overall conclusions are not peculiar to it. Our main takeaway points can be summarized as follows:
\begin{itemize}

\item The inferred precision $\sigma_{\fnl}$ depends very sensitively on the value of $p$ (cf.~Fig.~\ref{fig:pfixed} and Tab.~\ref{tab:pfixed}). Within the range $p \in \left[-1, 3\right]$ we explored, the constraints on $\sigma_{\fnl}$ displayed variations of over a factor of $\approx 14$. This strong sensitivity is as expected from the degenerate contributions of $\bphi$ and $\fnl$ to the galaxy power spectrum, and the fact that $\bphi(b_1)$ can be very close to zero for values of $p$ close to the linear density bias $b_1$ of the BOSS DR12 galaxies.

Although in analyses with fixed values of $p$ the significance of detection (SoD) of $\fnl \neq 0$ is not strongly affected by the choice of $p$ (cf.~Fig.~\ref{fig:pfixed}), the strong dependence of $\sigma_{\fnl}$ on $p$ still makes these analyses inadequate ways to quote constraints on local PNG (cf.~discussion in Sec.~\ref{sec:presults_pfixed}).

\item Marginalizing over the uncertain value of $p$ with wide uninformative priors is not conservative and biases the constraints through projection effects (cf.~Fig.~\ref{fig:ppriors} and Tab.~\ref{tab:pprior}). For Gaussian priors centered at $p_{\rm center} = 0.55$ and $p_{\rm center} = 1$, increasing the prior width from $\Delta p = 0$ to $\Delta p = 2$ results first in a gradual increase of the error bar on $\fnl$ up to factors of $\approx 5$, and then a subsequent decrease towards $\sigma_{\fnl} \to 0$ as larger values of $\Delta p$ progressively (and artificially) narrow down the constraints around $\fnl = 0$ (cf.~discussion in Sec.~\ref{sec:presults_pprior}).

\item Independently of $\bphi$ uncertainties the scale-dependent bias signature on the galaxy power spectrum can only be used to constrain the parameter combination $\fnl\bphi$, which can still be used to assess the SoD of $\fnl \neq 0$ (cf.~Fig.~\ref{fig:bphifnl}). For the BOSS DR12 galaxies we found an overall SoD of $1.6\sigma$, consistent with no detection. 

\end{itemize}

Overall, our results show that, until we develop a robust knowledge of the galaxy bias parameter $\bphi$ and its relation to $b_1$, {\it any observational constraints and forecasts on $\fnl$ using the scale-dependent bias effect are subject to a large theory systematic error and must therefore be interpreted carefully.} This strongly motivates revisiting the way we currently quote constraints on local PNG using galaxy data, in particular, that the constraining power of different galaxy surveys and analysis choices should be compared at the level of the SoD and the $\fnl\bphi$ parameters, and not $\fnl$ and $\sigma_{\fnl}$.

Our results encourage in particular the development of research programs dedicated to design accurate priors for the $\bphi(b_1)$ relation. This relation for dark matter halos in simulations is relatively well understood, although note it possesses a strong halo assembly bias signal \cite{2010JCAP...07..013R, 2022arXiv220907251L}. The situation for simulated galaxies is far less well understood. In Refs.~\cite{2020JCAP...12..013B, 2022JCAP...01..033B}, the authors took the first steps to study the $\bphi(b_1)$ relation with separate universe simulations of the IllustrisTNG model, and found significant differences to the halo-based relations. The sensitivity of these results to the assumed galaxy formation physics, as well as the connection between the simulated and the observed galaxy samples remains however currently unknown. Another interesting route could be to estimate the $\bphi(b_1)$ relation with the aid of semi-analytical models of galaxy formation tuned to match certain properties of galaxies in real surveys \cite{2022arXiv220411103A}. 

Should future research directions like these succeed in providing trustworthy priors on the PNG bias relations, this will also offer us a chance to optimize galaxy clustering analyses to $\fnl$ constraints by targeting galaxy samples with the largest expected values for $\bphi$. For example, Ref.~\cite{2022arXiv220907251L} finds that galaxies that preferentially inhabit higher-concentration halos tend to have larger values of $\bphi$. Further, the results in Ref.~\cite{2022JCAP...01..033B} suggest that objects with lower black hole mass accretion rate (or by proxy, lower black hole luminosity) could also be used to select objects with larger values for $\bphi$. The work of Ref.~\cite{2019MNRAS.483.4501A} showed also that splitting galaxies in SDSS by measures of their environment can have a strong impact on the resulting values of the $b_1$ parameter; using simulations, it would be interesting to investigate the impact of these splits on the $\bphi$ parameter as well. Note that if the goal is just to detect the parameter combination $\fnl\bphi$, then the accuracy requirements on these priors are actually not as stringent: in this case, even a rough understanding of which galaxy types are expected to have the largest values of $\bphi$ would be helpful to construct galaxy samples that provide higher chances to detect $\fnl \neq 0$. A rough understanding of at least the redshift evolution of the $\bphi$ parameter is useful also for analyses that employ optimal redshift weighting schemes to constrain local PNG \cite{2019MNRAS.485.4160M, 2019JCAP...09..010C}. However, the strongest motivation for these types of works is perhaps that, without very accurate and precise priors on the $\bphi(b_1)$ relation, it may remain impossible to constrain the actual numerical value of local $\fnl$ using the scale-dependent bias effect.

% ============================================================================================== %
% ============================================================================================== %
\acknowledgments
We would like to thank Emanuele Castorina, Eiichiro Komatsu and Fabian Schmidt for very useful comments and conversations. We are also very thankful to Oliver Philcox for making publicly available the BOSS DR12 power spectrum measurements utilized in this paper. The author acknowledges support from the Excellence Cluster ORIGINS which is funded by the Deutsche Forschungsgemeinschaft (DFG, German Research Foundation) under Germany's Excellence Strategy - EXC-2094-390783311. The numerical analysis presented in this work was done on the Cobra supercomputer at the Max Planck Computing and Data Facility (MPCDF) in Garching near Munich.

\appendix 
% ============================================================================================== %
% ============================================================================================== %
\section{Additional constraint plots}
\label{app:triangle}

\begin{figure}
\centering
\includegraphics[width=\textwidth]{\pathtofigs 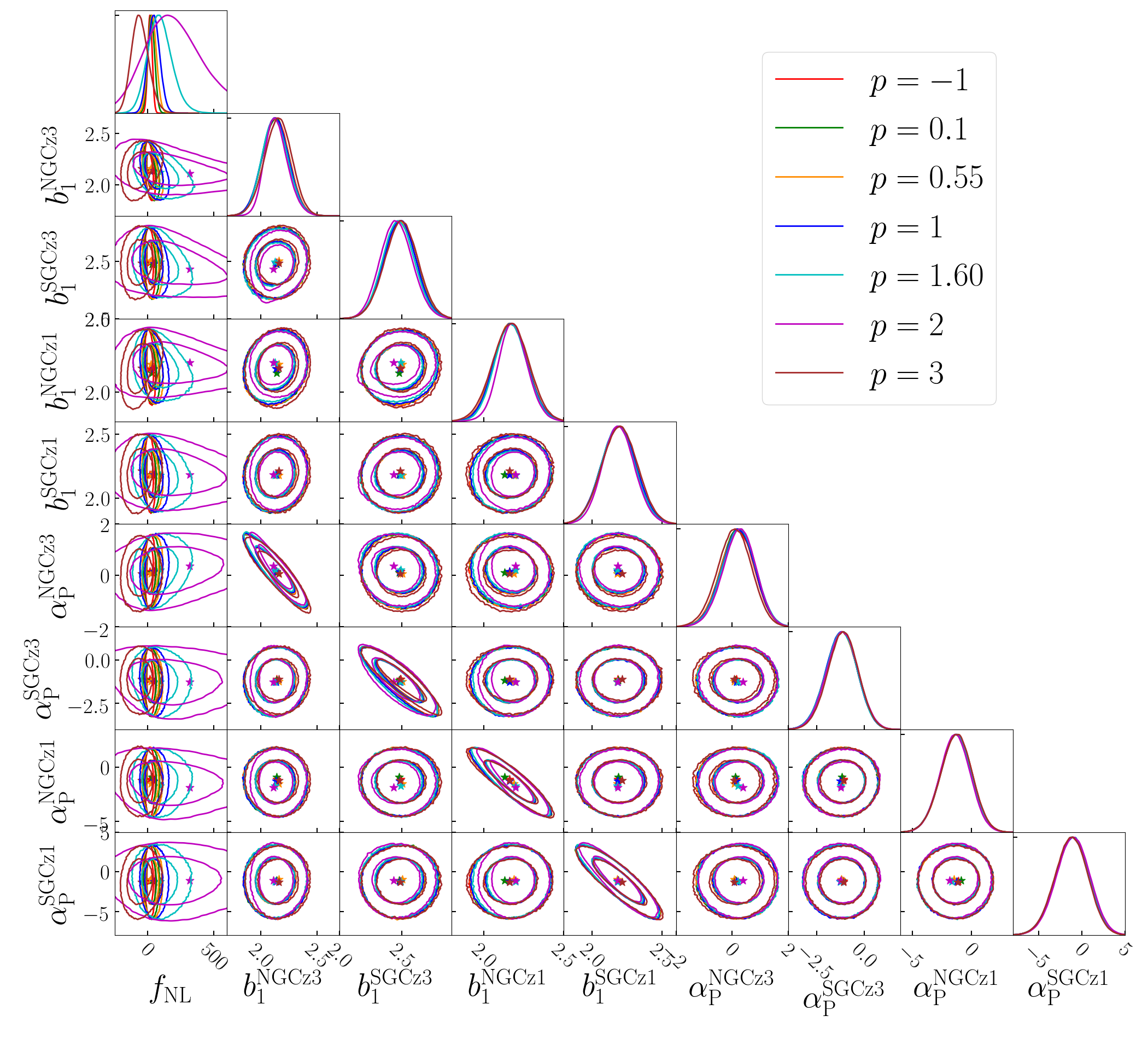}
\caption{Corner plot with one- and two-dimensional ($1\sigma$ and $2\sigma$) marginalized constraints from the analysis with different assumed fixed values of $p$ in the parametrization $\bphi(b_1) = 2\delta_c(b_1-p)$. This result is for Gaussian priors assumed on the parameter $b_1$, and corresponds to Fig.~\ref{fig:pfixed} in the main body of the paper.}
\label{fig:triangle_pfixed}
\end{figure}

\begin{figure}
\centering
\includegraphics[width=\textwidth]{\pathtofigs 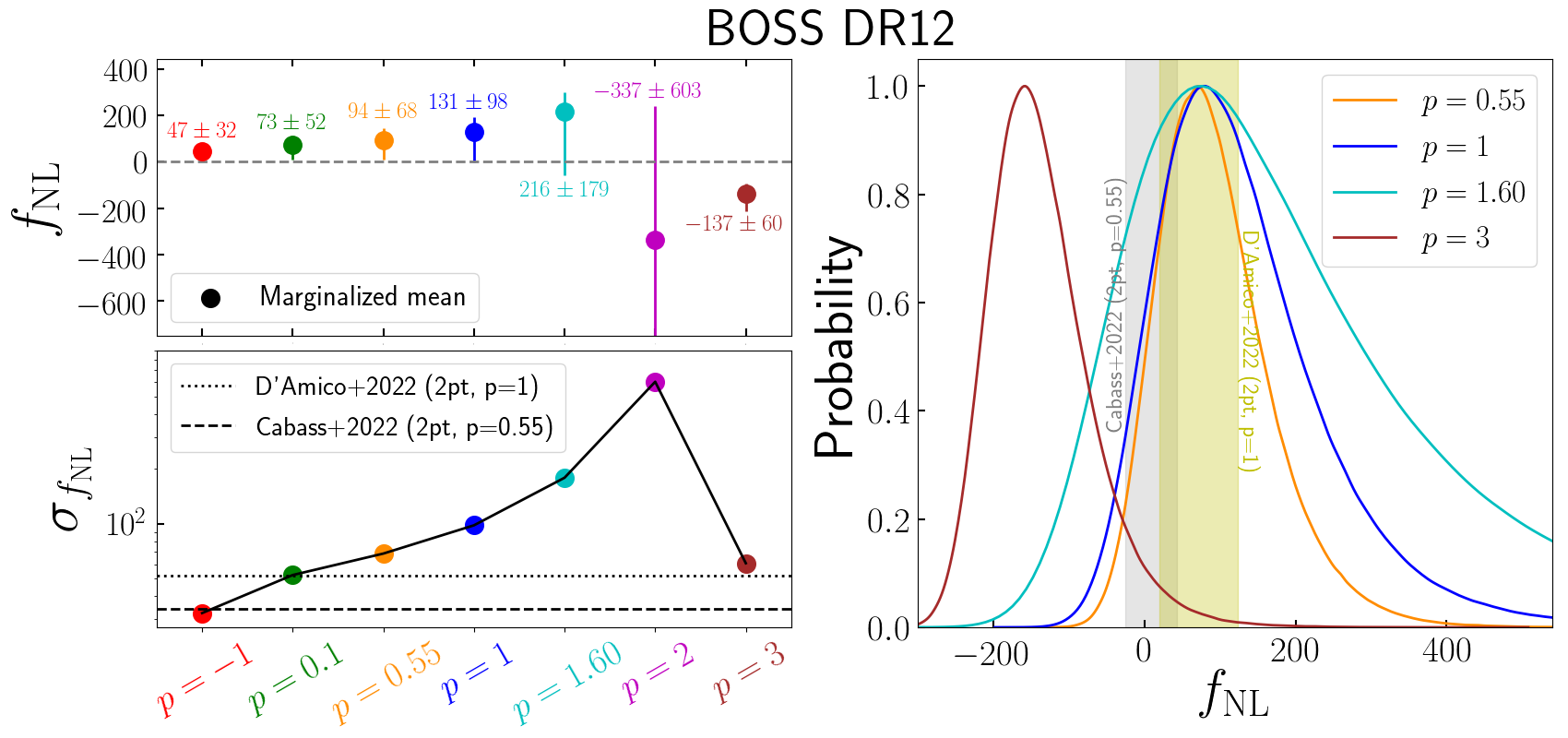}
\caption{Same as Fig.~\ref{fig:pfixed} in the main body of the paper, but without assumed Gaussian priors on the parameter $b_1$. The absolute value of the statistical error $\sigma_{\fnl}$ increases when $b_1$ is varied freely, but importantly, the main conclusions about the impact of different choices of $p$ and the $\bphi(b_1)$ relation remain unaffected.}
\label{fig:pfixed_nob1priors}
\end{figure}

\begin{figure}
\centering
\includegraphics[width=\textwidth]{\pathtofigs 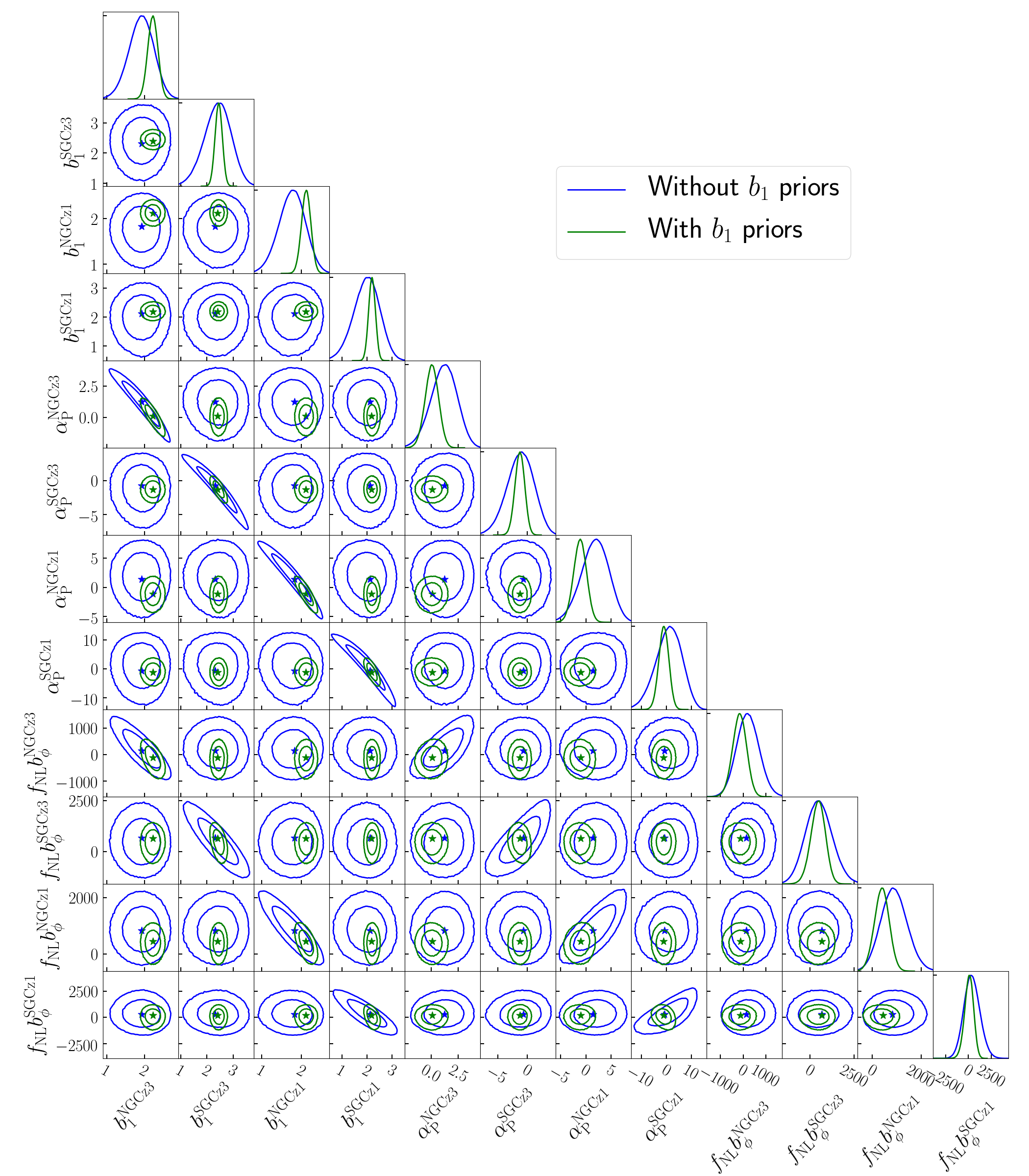}
\caption{Corner plot with one- and two-dimensional ($1\sigma$ and $2\sigma$) marginalized constraints from the analysis that fits for the parameter combinations $\fnl\bphi$. This corresponds to Fig.~\ref{fig:bphifnl} in the main body of the paper.}
\label{fig:triangle_fnlbphi}
\end{figure}

In this appendix, we display a few additional plots with one- and two-dimensional parameter constraints from parts of the analysis in the main body of the paper. Concretely, 

\begin{itemize}

\item Figure \ref{fig:triangle_pfixed} shows the corner constraints plot for all of the parameters varied in the fixed-$p$ analysis discussed in Sec.~\ref{sec:presults_pfixed}, for the case with Gaussian priors assumed on $b_1$.

\item Figure \ref{fig:pfixed_nob1priors} shows the same as Fig.~\ref{fig:pfixed}, but for the constraints without assumed priors on $b_1$.

\item Figure \ref{fig:triangle_fnlbphi} shows the corner constraints plot for all of the parameters varied in the analysis of the SoD parameters $\fnl\bphi$ in Sec.~\ref{sec:sod}.

\end{itemize}

% ============================================================================================== %
% ============================================================================================== %
\bibliographystyle{utphys}
\bibliography{REFS}

\end{document}